\documentclass[letterpaper,twocolumn,10pt,hyphens]{article}
\usepackage{usenix2019_v3}

\newif\ifdraft
\drafttrue
\draftfalse

\newif\ifspace
\spacetrue
\spacefalse

\usepackage{booktabs} 
\usepackage{graphicx}
\usepackage{comment}
\usepackage{amsmath,amssymb,amsfonts}
\usepackage{wrapfig,paralist}
\usepackage{lscape}
\usepackage{rotating}
\usepackage{epstopdf}
\usepackage{color}
\usepackage{amsthm}
\usepackage[boxruled,linesnumbered]{algorithm2e}
\usepackage{hyperref}
\usepackage[capitalise, noabbrev]{cleveref}
\usepackage{savesym}
\savesymbol{state}
\usepackage{tengwarscript}
\usepackage{xcolor}
\usepackage[]{url}

\usepackage[noend]{algpseudocode}
\usepackage{array}
\usepackage{enumitem}
\usepackage{hyperref}
\usepackage{xspace}
\usepackage{multirow}
\usepackage{multicol}
\usepackage{subcaption}
\usepackage{pdflscape}
\usepackage{mathtools}
\usepackage{booktabs}
\usepackage{float}
\usepackage{newfloat}
\usepackage{tikz}
\usepackage{adjustbox}
\usepackage[utf8]{inputenc}
\usepackage{array}
\usepackage{makecell}

\usepackage{tabularx}
\usepackage{caption}
\usepackage{stfloats}
\usepackage{arydshln}

\usepackage{colortbl}

\SetCommentSty{mycommfont}
\usepackage{ulem}

\newcommand{\oldtodo}[1]{}

\newcommand{\oldtext}[1]{}

\ifdraft

\newcommand{\newtext}[1]{\ifdraft\textcolor{blue}{{#1}}\fi}
\newcommand{\newtextMM}[1]{\ifdraft\textcolor{blue}{{#1}}\fi}

\newcommand{\prerevtext}[1]{{\sout{#1}}}
\else

\newcommand{\prerevtext}[1]{}

\newcommand{\newtext}[1]{#1}
\newcommand{\newtextMM}[1]{#1}
\fi

\renewcommand\paragraph[1]{\vspace{0.3mm}\noindent \textbf{#1.\ }}

\newcommand{\figurescale}{0.5}
\newcommand{\figurewidth}{0.9}

 \definecolor{mygray}{gray}{0.6}
\newcommand{\snote}[1]{\textcolor{mygray}{#1}}

\begin{document}
\pagestyle{plain}

\title{Empirical Understanding of Deletion Privacy\\ Experiences, Expectations, and Measures}
\author{
	Mohsen Minaei\\
	Purdue University\\
	mominaei@visa.com
	\and
	Mainack Mondal\\
	IIT Kharagpur\\
	mainack@cse.iitkgp.ac.in
	\and
	Aniket Kate\\
	Purdue University\\
	aniket@purdue.edu
}

\maketitle
\begin{abstract}

\noindent In recent years, social platforms are heavily used by individuals to share their thoughts and personal information. 
However, due to regret over time about posting inappropriate social content, embarrassment, or even life or relationship changes, some past posts might also pose serious privacy concerns for them.
To cope with these privacy concerns, 
social platforms offer deletion mechanisms that allow users to remove their contents.
Quite naturally, these deletion mechanisms are really useful for removing past posts as and when needed. However, these same mechanisms also leave the users \textit{potentially} vulnerable to attacks by  adversaries who specifically seek the users' damaging content and exploit the act of deletion as a strong signal for identifying such content. Unfortunately, today  user experiences and contextual expectations regarding such attacks on deletion privacy and deletion privacy in general are not well understood.

To that end, in this paper, we conduct a user survey-based exploration  involving 191 participants to unpack their prior deletion experiences, their expectations of deletion privacy, and how effective they find the current deletion mechanisms.
We find that more than 80\% of the users have deleted at least a social media post, and users self-reported that, on average, around 35\% of their deletions happened after a week of posting.
While the participants identified the irrelevancy (due to time passing) as the main reason for content removal, most of them believed that deletions indicate that the deleted content includes some damaging information to the owner. Importantly, the participants are significantly more concerned about their deletions being noticed by large-scale data collectors (e.g., a third-party data collecting company or the government) than individuals from their social circle. 
Finally, the participants felt that popular deletion mechanisms, although very useful to help remove the content in multiple scenarios, \newtext{are} not very effective in protecting the privacy of those deletions. Consequently, they identify design guidelines for improving future deletion mechanisms.
 
\end{abstract}

\section{Introduction}




\noindent Today, billions of internet users share hundreds of billions of pieces of their personal content (including life events, images, and opinions) on social platforms like Facebook or Twitter. A recent Pew Research study~\cite{pewresearch18socialusage} finds that seven out of ten American adults use some kind of social platform. As these platforms archive a significant number of posts over the years, 
the often-personal nature of this social content does make the users uncomfortable at times.
Such discomfort may originate from regret over time about posting inappropriate social content, embarrassment, or even life or relationship changes~\cite{tweets-forever,wang2011regretted,zhou-2016-tweetproperly,bauer-2013-postanachronism,ayalon-2013-retrospective}. 


Almost all social platforms offer some mechanism to their users to retrospectively remove their unwanted posts, and \newtext{selective deletion is frequently used}---\newtext{around 5\% of posts that are more than six years older were removed} from Twitter by 2016~\cite{mondal2017longitudinal,mondal-2016-longitudinal-exposure}. However, these selective deletions create a catch-22 situation---deletions (meant to remove a post from the platform), in practice, might bring unwanted attention to deleted posts and make them more visible to an onlooker. 
Many web services (e.g., Politwoops~\cite{Politwoops} for Twitter, Removeddit~\cite{removeddit} for Reddit, StackPrinter-Deleted~\cite{StackPrinter} for Stack overflow, and YouTomb~\cite{YouTomb}) collect and hoard specific deleted posts from 
social platforms. Such services are intuitively expected---due to the open and social nature of these platforms, it is relatively easy to collect snapshots and  identify deleted social content. However, this social content (by virtue of being selectively deleted) might already contain embarrassing or potentially damaging information. 
Indeed, a line of work identifies the phenomena of leveraging deletion as a signal to unearth potentially sensitive content as a violation of ``deletion privacy''~\cite{minaei2019lethe,minaei2020deceptivendss,wang2019donttweetthis,tweets-forever,zhou-2016-tweetproperly}, and 
there are already several deployed as well as academic removal mechanisms devoted to preserving deletion privacy~\cite{snapchat-site,ig-stories,4chan,minaei2019lethe,minaei2020deceptivendss}. 

From analyzing these mechanisms, we observe that preserving deletion privacy involves designing and enforcing access control rules to regulate when and how information about the deletion events (and deleted content) is revealed to others. However, earlier research did not attempt to uncover these systematic access rules that regulate the desired discoverability of deletion events.
More specifically, there is no prior work on understanding the \textit{need} for providing deletion privacy to social media users. In other words, there is no evidence quantifying the importance of preserving deletion privacy for social platform users. To that end, this paper takes the first step towards understanding and operationalizing user perceptions of deletion privacy. 

Specifically, in this work, we conducted a survey-based user study to uncover the need as well as contextual access control norms governing deletion privacy in social platforms. Our survey, collected quantitative and qualitative data from 191 participants spanning Europe and the US regarding their perceptions of deletion privacy. We first investigated the prior experiences of the participants regarding their post deletions and corresponding deletion privacy expectations. We then leverage contextual integrity theory~\cite{nissenbaum-2010-ci-book} to identify key contextual factors (as perceived by users) for regulating access and ensuring the preservation of deletion privacy. Finally, we unearth the factors governing the usefulness of existing deletion 
mechanisms. Specifically, we investigate the following research questions (RQ). 


\vspace{1mm}

\textbf{RQ1:} \textit{What are users' experiences with deletions in social platforms? \newtext{Did some other users or organizations focus on (or notice) their deleted social posts?} How?}

We investigated this RQ by asking each participant detailed questions regarding their experience with deletions on social platforms. We note that 82\% of our participants  deleted some of their posts due to a number of reasons. This huge fraction indicates the wide usage and consequent utility of deletions to users of social platforms. Interestingly, 51\% of the participants felt that a deleted post is \textit{sensitive, damaging, or embarrassing} to its owner. Furthermore, 54 participants were aware of others noticing their post deletions in social platforms, \newtext{and within our sample of fewer than 200 participants, nine participants pointed out that when others noticed their deletions, it resulted in discomfort.} 

While establishing the need for deletion privacy was a prime goal of this work, a social platform would also need to know if its users feel (un)comfortable in revealing their deletions in a certain context. We explore this question next. 

\vspace{1mm}
\textbf{RQ2:} \textit{On what contextual factors 
do rules regarding the acceptability of revealing deletion events depend? How?} 

We used \newtext{contextual integrity theory}~\cite{nissenbaum-2010-ci-book} to create a set of contextual variables (e.g., recipients) and enumerated possible values for each set of variables (e.g., family member, friend, coworker, a company, government). We then collected user feedback for combinations of all of those contextual variables in our survey. We observe that most users seek to preserve deletion privacy against large-scale data collectors (e.g., corporations and government), but not so much against their family, friends, and even coworkers.

Finally, we looked into the efficacy of the existing deletion-privacy-enhancing mechanisms. To that end, we asked:

\vspace{1mm}
\textbf{RQ3:} \textit{Are existing mechanisms effective and useful for enhancing deletion privacy? Why or why not?} 

We based this part of our study using short videos that explained the high-level functionalities of different deletion mechanisms. These mechanisms provide varying guarantees to protect deletion privacy. Users find selective deletions (the current deletion mechanism used by most social platforms) to be ineffective in protecting their sensitive deletions. The same users found the other mechanisms more effective (they  also identified their shortcomings). We provide a principled analysis of the pros and cons of each of the existing mechanisms via examining the users' perceptions.

In summary, we begin to quantify  the need for deletion privacy in social platforms with a 191 participant user survey in this work. Our key contributions include:

    1. Establishing the users need to ensure deletion privacy in social platforms. Our results demonstrate that users widely leverage available deletion mechanisms, but they also care about protecting their deletion privacy.
    
    2. Showing the context-dependency of the rules for preserving deletion privacy. We identify the key contextual factors that future developers should consider to align their system functionalities with user expectations better. 

    3. Identifying key factors governing the  usefulness of deletion mechanisms to preserve deletion privacy. Future social platform designers should consider these factors to ensure the deletion privacy of users.

\section{Background and Related Work}

\subsection{Deletion in Social Platforms}\label{sec:deletion_media}




\noindent Deletion or the ability to remove content is a crucial functionality in social platforms---often needed due to the social nature of these platforms and the personal nature of social content. Earlier work studied in detail the reasons behind social content deletion. These reasons range from removing regrettable content to removing content, which became irrelevant over time~\cite{tweets-forever,wang2011regretted,zhou-2016-tweetproperly,mondal-2016-longitudinal-exposure,mondal-2019-retrospective-facebook, schnitzler2021sok}. 
In fact, multiple prior works aimed to unpack user perceptions regarding their understanding of deletion mechanisms and the impact of deletions on post owners and other users. For example, Murillo et al. found that users did not clearly understand that deletion on their social platform interface might not guarantee the removal of data from other places (e.g., the platform's servers or even other users' walls)~\cite{murillo-deletion-2028-soups}. 
Another recent work found that some users want to get notified if the content is deleted by original owners~\cite{yilmaz-perceptions-2021-icwsm}. These results underline the opaqueness of current deletion mechanisms (leading to misunderstanding) and the willingness of users to check for others' deletions in specific social contexts. Consequently, these earlier works further motivate a need to investigate if users perceive that deletions, while extremely useful, might also create novel privacy problems. We address this need in this work.

Specifically, even though deletion is crucial and widely adopted, in some cases, the removal of a post can be a simple and very effective indication of sensitive and/or damaging social content~\cite{minaei2019lethe,minaei2020deceptivendss} in the post. \newtext{Thus, a simple removal \textit{might create} an opportunity for an attacker to harass and blackmail the users.} Such deletion-based surveillance is not only relevant to public figures, but also, for normal social platform users, e.g., a French Twitter account, @FallaitPasSuppr, identifies and re-publishes deleted content of both French public figures as well as normal users~\cite{fallait}. This situation is akin to \textit{censorship backfire} in the Streisand case, where censoring (e.g., via removal) content actually attracts more attention and causes a privacy violation~\cite{jansen_streisand_2015}. Related works on understanding censorship from the legal domain are highly relevant yet orthogonal to our study---our efforts focus on primarily understanding the user perceptions as well as access control rules regarding deletion privacy violation which can be utilized in computational system design. Aside from censorship backfire, our work is also built on the discourse on the ``Right to be Forgotten'' article in GDPR~\cite{GDPR_art17-right-to-be-forgotten}. Specifically for European users, a deletion privacy violation can also be viewed as a violation of the   ``Right to be Forgotten'' (where a user wants everyone to forget a deleted post). However, so far the ``Right to be Forgotten'' and perceptions regarding this right is never explored in this context of deletion privacy in social platforms. Our work is one step in that direction.

In this study, our participants' self-reported social content deletion behavior is quite in line with earlier work (both qualitative and quantitative). E.g., in our study, 24\% of the post deletions happened within less than two minutes from posting time (similar to~\cite{tweets-forever}, and 35\% of tweets were deleted in the long term. 
Many of the self-reported reasons behind deletions in our study are also aligned with prior work (e.g., Fixing Spelling or Grammar). However, unlike some earlier studies, a majority of our post deletions happened due to ``Being irrelevant due to time passing,'' hinting at the correlation between post-withdrawal and time~\cite{ayalon-2017-past}. 
However, these prior works did not investigate the extent and potential impact of adversaries finding out deletion events, i.e., they did not explore \text{deletion privacy}. In this work, we fill this gap and show (using self-reported data) that violation of deletion privacy is indeed a very real problem for our participants. \newtext{Furthermore, our study identified concrete norms governing the privacy of deletion events.}



\subsection{Social Content Deletion Mechanisms}\label{sec:mechanism_background}

\noindent We identify four key mechanisms for facilitating deletion in social platforms: 

\noindent\textbf{Selective Deletion:} The Majority of the social platforms today provide a selective deletion mechanism--the posts are available on the platform until the users themselves select an unwanted post and delete it. However, 
selective deletion might attract unwanted attention to particular posts~\cite{minaei2019lethe}. 

\noindent\textbf{Prescheduled Deletion:} This mechanism automatically removes the users' contents when a specific criterion has been triggered (e.g., after a  predefined period or after prolonged inactivity around the post~\cite{mondal2017longitudinal}). E.g., Snapchat and Instagram Stories support this feature where they delete posts after some time. This mechanism ensures that an adversary cannot single out the damaging deleted content as all posts are deleted. However, it removes \textit{everything} on the downside, implying 
no archive of social content for users to reminiscence. 

\textbf{Intermittent Withdrawal:} Intermittent Withdrawal~\cite{minaei2019lethe} offers a deniability guarantee for users' deletions using an availability-privacy tradeoff.
In this mechanism, all of the non-deleted posts are intermittently hidden for some amount of time. This hiding confuses an adversary while deciding if an unavailable post is deleted by the user or temporarily hidden by the platform.

\textbf{Decoy Deletions:} In Decoy Deletions~\cite{minaei2020deceptivendss}, given a set of damaging posts that users want to delete, the system selects $k$ additional non-damaging posts for each damaging post and deletes them along with the damaging posts. The system-selected posts (decoy posts) are taken from a pool of non-damaging, non-deleted posts provided by volunteers. Decoy Deletions raises the bar for the adversary to identify deleted posts as they need to identify the sensitive or damaging post among the $k+1$ deleted posts. 

Although some of these deletion mechanisms aimed to facilitate private deletion in social platforms, they did not consider a fundamental question: is preserving deletion privacy at all important for end-users today? In this work, we answered this question affirmatively, and our study compares the effectiveness of these mechanisms to protect deletion privacy.

\begin{table*}[t]
\vspace{0mm}
	\caption{{Contextual integrity (CI) parameter values used to generate information flow of deletion events}\vspace{-2mm}}
	\centering
	\resizebox{0.99\textwidth}{!}{
		\begin{tabular}{|l|l|l|}
			\hline
			\textbf{Sender}             & \textbf{Transmission Principle}                                       &   \textbf{Attributes}                         \\
	        user itself                 &  because they were checking/observing your user profile regularly     &   Post did not get enough attention           \\
		    \cline{1-1}
		    \textbf{Recipient}          &  because they were mentioned in the post or interacted with the post  &   Fixing Spelling/Grammar                     \\
			your family member          &   \textit{null}                                                       &   Cleaning up profile for new job             \\
			                            \cline{2-2}
			your friend                 &   \textbf{Subject}                                                    &   Cleaning up profile for new relationship    \\
			your coworker/acquaintance  &   that contained some information about yourself                      &   Racial/Religious/Political reason           \\
			a company                   &   that contained some information about your family members           &   Being irrelevant due to time passing        \\
			the government              &   that contained some information about your friends                  &   Removing sexual content                     \\
			anyone                      &   that contained some information about your coworkers                &   Removing drug/alcohol related content       \\
			                            &   \textit{null}                                                       &   Removing violence/cursing related content   \\
			                            &                                                                       &   Removing health related content             \\
		    \hline
		\end{tabular}
	}
	\label{tab:ci_infoflow}
\end{table*}

\subsection{Deletion Privacy as Contextual Integrity} \label{sec:ci_background}
\noindent Contextual Integrity (CI)~\cite{nissenbaum-2010-ci-book} theory provides a systematic framework for studying privacy norms and expectations. CI defines privacy as appropriate flows of information. Each information flow consists of five parameters: subject, sender, recipient, information type (or attribute), and transmission principle. The appropriate information flows conform to the socially acceptable values of these parameters.
Earlier work demonstrates that we can infer privacy norms (i.e., rules regulating acceptable information flow) by measuring the acceptability of different information flows (created with a varying combination of CI parameter values)~\cite{apthorpe2018discovering, apthorpe2019evaluating}.
For example, users in general might be comfortable when a fitness tracker (\textit{sender}) sends user's heart rate (\textit{attribute}) to the doctor (\textit{recipient}) to monitor the health status (\textit{transmission principle}), but uncomfortable if the \textit{recipient} is a health insurance provider. Given, a primary aim of this work is to understand societal rules of deletion privacy we choose CI as a suitable framework to unpack these rules. Consequently, we build our work on the existing body of literature on CI and its applications.

Here, we aim to discover the effect of context on the acceptability of deletion events getting noticed. 
Thus, we modeled this problem (deletion privacy) as simply ensuring the  appropriateness of the flow of a deletion event initiated by a user and noticed by a receiver.

Our work focuses on the flow of the deletion event information, which starts from the user (when (s)he initiates content deletion) to the receiver (who notices the deletion). In this set-up for each flow, the \textit{sender} is the user (who deleted), \textit{subject} is who the deleted post was about (might be some other groups of users).

We selected the CI parameter values relevant to deletion privacy by surveying earlier work and conducting pilot studies. \cref{tab:ci_infoflow} contains the full list of our CI parameter values.
Note that this list is not exhaustive. However, as a first, it does cover a range of information flows in the scope of deletion privacy and demonstrates the generality of our approach. \cref{sec:survey_instrument} details the exact questions asked in our survey. Next, we present our CI parameter values.


\textbf{Sender \& Subjects}
In this set-up for each flow, the sender is the user (who deleted), and the subject is who the deleted post was about.
We considered prior work on ego networks and social circles to design four distinct subjects ~\cite{hill-socialnetsize-03,arnaboldi-2012-activityEgonet}---(i) the user, (ii) family members, (iii) friends, (iv) coworkers/acquaintance. We also included ``not specifying a subject'' as \textit{null} i.e., control condition.

\textbf{Recipients}
 for social content deletion is the individual (or organization) that notices the user's deletion. We include users' social circles in our list of recipients along with two other entities--a company that collects and archives users' deletions 
and the government.

\textbf{Transmission Principles}
 are methods that a recipient uses to discover the deletion. We considered three discovery methods---(i) discovery by checking/observing the user profile regularly to observe any user-profile change 
(ii) discovery due to an interaction with the post (e.g., liking, commenting, reposting, sharing, etc.), 
(iii) not specifying a discovery method (\textit{null} i.e., control condition).

\textbf{Attributes},
we consider the reason of the deletion to be the attribute in the information flows. 
We adapt the categories defined by Zhou et al.~\cite{zhou-2016-tweetproperly} for the regrettable deleted tweets as attributes. We further add ``fixing spelling/grammar'' from earlier work~\cite{almuhimedi2013tweets} as well as two other reasons---``post did not get enough attention'' and ``being irrelevant due to time passing'' based on our study pilot.
We obtained feedback on acceptability for the information flows generated using combination of all of these CI parameter values. 

\section{Methodology}
In this section, we discuss the design of our study in detail.
We begin with our survey instrument that paves the path for understanding the users' experience with deletion events, unrolling the deletion privacy norms using Contextual Integrity (CI), and evaluating the effectiveness of deletion mechanisms in providing privacy to the deletions.

\subsection{Survey Instrument}\label{sec:survey_instrument}
\noindent Our survey instrument was constructed of two parts. 
In the first part, we asked questions about the users' past experiences of deletions on social platforms and later asked about their deletion privacy preferences using CI.
In the second part, we probed the effectiveness and usefulness of different deletion mechanisms in hiding the users' unwanted content.
Our full survey instrument 
can be accessed in
\cref{sec:survey_full}.

\subsubsection{Part I: Perceptions of Content Deletion}
Part I of our survey contained two sections: (1) Experiences about prior post deletions, (2) CI-parameter based questionnaire about deletion privacy.

\paragraph{Experiences about prior post deletions (RQ1)}
We started by asking the participants about their usage of different social platforms and whether they have ever deleted any of their content. 
We further asked how old the content was at the time of deletion and the reasons behind their deletions.
We then inquired if the participants were aware of other users noticing their deletions and whether they have noticed other users' deletions. 
Lastly, we investigated how the users felt about the sensitivity of deleted content by asking them whether they agree or disagree with the statement---``when someone deletes a social media post, it indicates that the content of that post is sensitive/damaging/embarrassing to that individual.''

\paragraph{CI-based questionnaire: deletion privacy (RQ2)}
Taking inspiration from earlier research~\cite{apthorpe2018discovering,apthorpe2019evaluating}, 
we adapted a CI-based questionnaire to investigate the users' expectations of deletion privacy in social platforms. 
We detail our adaptation of CI to the scope of deletion privacy in~\cref{sec:ci_background}. Here, we will focus on the setup of our survey. 

We needed to obtain users' perception of acceptability for the information flows created by all combinations of the parameter values presented in~\cref{tab:ci_infoflow}. 
In total, we had 900 distinct information flows (5 \textit{subjects} $\times$ 3 \textit{transmission principles} $\times$ 6 \textit{recipients} $\times$ 10 \textit{attributes}), and asking each participant to evaluate all the flows was infeasible. Therefore, we divided the flows into 30 blocks with 30 information flows each, where every block was assigned to at least 6 participants.

To randomly assign participants to one of these blocks,  
each participant was  randomly assigned to a \textbf{fixed} value for the \textit{subject} and \textit{transmission principle} variable. 
That participant was also randomly assigned to one of the two pre-defined sets of \textit{recipient} variable values\footnote{
	First set of \textit{recipient} variable values or recipient\_A: [your family member, your close friend, your coworker].
	Second set of \textit{recipient} variable values or recipient\_B:[anyone, a company, the government]} 
(each set contained three \textit{recipient} values).

As a result, in each block, we repeated the below matrix question three times by replacing the \textit{recipient} variable with the values from
the assigned recipient set, 
but keeping the same \textit{subject} and \textit{transmission principle} values each time.
The rows of this matrix question represented the \textit{attribute} variable values (an \textit{attribute} value represents the reason behind a deletion).
Therefore, each row of a question signified one of the information flows, which the participants were asked to rate its acceptability using a five-point Likert scale:
Completely Acceptable, Somewhat Acceptable, Neutral, Somewhat Unacceptable, Completely Unacceptable.

\textit{``CI-Q: We are putting a few possible reasons behind post deletions in the table below. 
	Imagine a situation where you deleted a post [subject] from one of your social media accounts due to that
	reason. In each of these situations, please indicate how acceptable is it for you that [recipient] notices your deletion [transmission principle]?''  
}

\subsubsection{Part II: Efficacy of Deletion Mechanisms (RQ3)}\label{sec:adversarial_model}
In the second part of the survey, we captured the effectiveness and usefulness of different deletion mechanisms from the perspective of the users.
We realized that this part involved possible hypothetical scenarios, as some participants may never have used some of the mechanisms. Thus, we took a visual (video) driven approach to first educate the users on these mechanisms and later ask about their efficacy. This approach is similar to those used in prior work on familiarizing participants with novel authentication mechanisms~\cite{lyastani-oakland2020-fido2}.  

To have a fair comparison between the different mechanisms, prior to introducing the mechanisms, we needed to present the threat model that we are considering protection against. 
In this work, we adapted the same threat model considered in earlier works on deletion privacy
~\cite{minaei2020deceptivendss,minaei2019lethe}. 
---\textit{In this threat model, the adversary can observe the entire social platform. Therefore, it can continuously access the platform to take snapshots of the posts with the goal of identifying the damaging/sensitive deleted posts to use against the users. The adversary aims to find as many damaging/sensitive posts as possible and does not perform targeted attacks on particular users.}
We demonstrated this threat model using a short video at the beginning of this section and asked the participants whether they have encountered attacks from such a malicious entity or how vulnerable do they find themselves against it.


\subsubsection{Quality control}\label{sec:attention_check}

To ensure the quality of responses, we incorporated multiple attention check questions in the survey. 
In particular, we repeated two of the multiple-choice questions in random locations in the survey and compared the answers with their previous responses.
 Moreover, we added a fake social platform named ``Cybersocial'' in questions that asked about their usage of social platforms to monitor whether they indicate using this platform or not. 
Further,
	we put two checks to ensure that participants indeed viewed the videos (and gave quality responses)---(i) we noted the view count of our YouTube videos periodically to ensure participant-viewing the videos (the videos were unlisted, thus a small possibility of random users viewing them) (ii) we reached out to participants whose total completion time was less than or very close to the total video-length and asked them to retake the survey. Some of them directly mentioned quotes from the video to demonstrate their understanding.

\subsection{Pilot Studies}
Prior to the survey's deployment, we conducted two pilot studies to evaluate the study's procedure, determine the average duration, and test the comprehensibility of the questions. 

In the first pilot, we tested the study on ten colleagues (without prior knowledge of the study and its goals) from different departments in our university.
As a result, we removed four questions from the survey as they were somewhat redundant or too imprecise. 
Moreover, some of the questions' choices were modified to eliminate any invasiveness and confusion. 

The major change 
was the reconstruction of the second part of the survey (i.e., the mechanisms' efficacy). Initially, we provided text descriptions of the threat model and 
the deletion mechanisms. However, the participants found the text descriptions to be monotonous, long, and, more importantly, hard to understand.
As a result, we modified this part 
and created video explanations of the deletion mechanisms (similar to~\cite{lyastani-oakland2020-fido2}). 
This allowed us to provide more information and present the mechanisms via examples and animations.

After applying the changes above, in the second stage of the pilot, we deployed the survey on the Prolific Academic~\cite{prolific} and recruited ten participants.
The results from the qualitative responses showed that the first pilot's changes were effective, and participants had a good understanding of the questions.
This point was also confirmed by the participant's responses to the following question at the end of the pilot survey:
``Did you find the questions in this survey to be understandable?''.
Eight of the participants responded with ``completely understandable'' and ``mostly understandable.''
The remaining two responded with ``neutral'' and ``mostly not understandable''; however, without any feedback on which sections they had difficulty in understanding.

\subsection{Recruitment}
We recruited our participants from Prolific Academic, 
a platform regularly used for advertising academic surveys~\cite{peer2017beyond}. 
We chose participants both from the US and Europe\footnote{Over 90\% of the participants in Prolific (as well as 80\% of participants in Amazon Mechanical Turk, AMT~\cite{amt}, another well known crowd-sourcing platform) are from the US and Europe~\cite{prolificdemographics,difallah2018demographics,pewamtdemographic}.}. 
We screened participants to ensure they were 18 years old or above, 
had not taken our pilot study, 
had taken a minimum of 50 prior surveys on the platform,  
had a minimum approval rate of 95\%, 
fluency in English,
and having a social media account currently or in the past.

While designing our survey instrument, we aimed to minimize bias (i.e., leading or priming) and ambiguity. We did not screen our participants based on deletion behavior and designed our recruitment text and strategy accordingly (presented in~\cref{sec:recruitment_message}). 
We carefully avoided priming participants by not using words like ``security'' or ``privacy'' in our study.
In the recruitment text, we used the keyword ``deletion'' only for describing a sub-part of the survey rather than a requirement. In fact, our reported percentages of participants who deleted posts are similar to the previous studies~\cite{mondal-2016-longitudinal-exposure,tinati2017instacan}.

The survey was advertised as ``A study about social media usage and post deletions''
and deployed in same sized batches (i.e., 20 participants at a time) over a one-week period, at different times of the day.   
We did this to counter anomalous time dependency in our results due to the effect of events happening at a specific time~\cite{albakry2020url}.
The average time of completion for each part of the survey was 12.5 minutes, and compensation was \$1.5 for each part (participants that completed both parts were compensated \$3). 
To gain some confidence in a fair payment, during our pilot study, we asked the participants---``How fair do you find the compensation of the survey, compared to the amount of time you took for completion?''.
90\% of the participants responded with ``very fair'' and ``fair''.
Furthermore, the median completion time was 22 minutes, resulting in compensation of \$8/hr, comparable to similar studies~\cite{yaqub2020effects}.

In total, we obtained 205 responses (103 from the US and 102 from Europe) for part I and 144 responses (93 from the US and 51 from Europe) for part II.

\subsection{Participant Demographics}
A total of 205 and 144 participants  completed parts I and II, respectively. 
We discarded the responses that did not pass the validity checks (see \cref{sec:attention_check}) and were left with 191 (93 from the US and 98 from Europe) and 135 (85 from the US and 50 from Europe) participants for parts I and II.
Our European participants were from 13 different countries where, 57\% were from UK, 13\% from Portugal, 11\% from Poland, and the remaining 19\% were from Hungry, Spain, Greece, Germany, Czech Republic, Netherlands, Sweden, Switzerland, Ireland, and Finland.       

Our population sample was nearly gender-balanced; 50.8\% identified as female, 47.6\% as male, and 1.1\% as others. 
The sample skewed young, with 26.7\% between 18 and 24, 
39.3\% between 25 and 34, 20.4\% between 35 and 44, and 13.6\% age 45 or older. 
Our participants were slightly more educated than the general U.S. population~\cite{educationattainment19}, where 55\% of the participants either had a bachelor or a graduate degree. 
The participants' median annual household income was reported as \$40,000 - \$59,999, where the majority had an income of \$20,000 - \$39,999 (23\%).
Even though participants in crowdsourcing platforms (e.g., Amazon Mechanical Turk and Prolific) are considered to be tech-savvy~\cite{hitlin2016turkers}, 67\% of our participants reported that they do not have any background (e.g., study, work, etc.) experience in the IT field. 
We present the detailed demographics of our population in
\cref{tab:demographics}.

\begin{figure}
	\centering
	\resizebox{\columnwidth}{!}{
	    \includegraphics{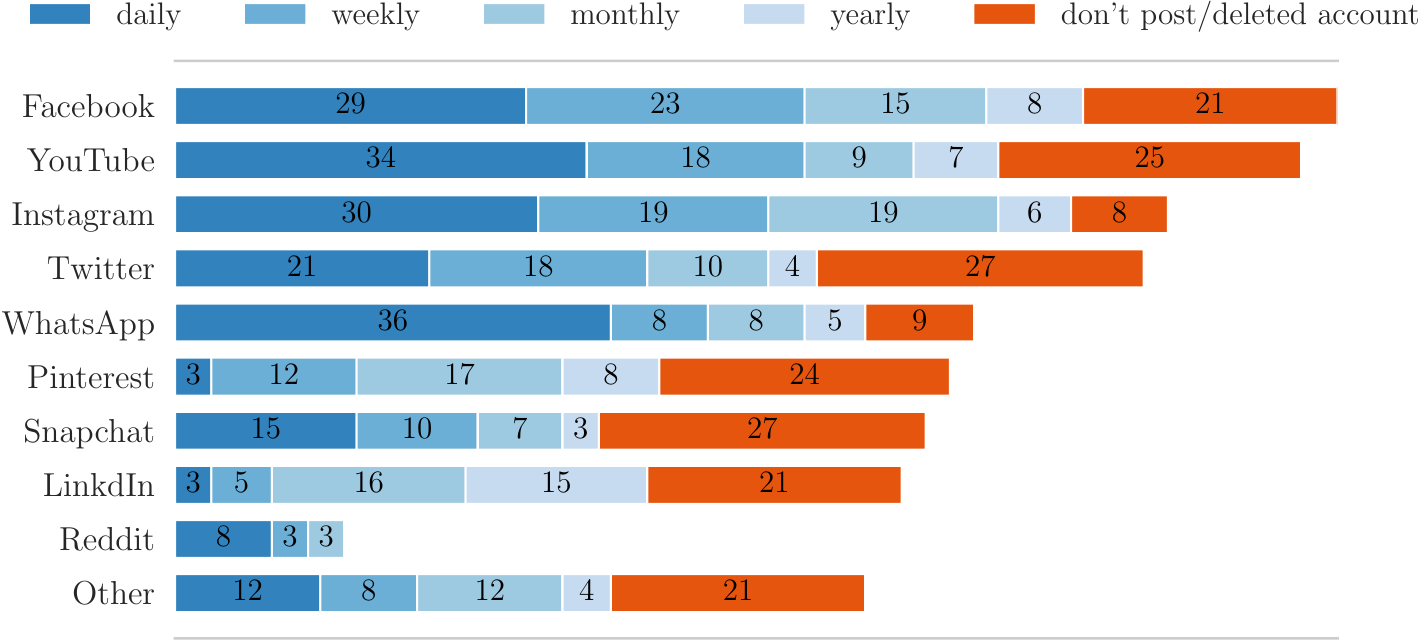}
	}
	\caption{The usage pattern (in percentage) of different social platforms by the participants.}
	\label{fig:platform_usage}
\end{figure}

Further, our participants are active users of popular social media platforms. 
74.9\% of the participants reported using at least one social platform daily, and 91.1\% use a platform at least once a week,
showing the participants' suitability for this study. 
Facebook, Youtube, Instagram, WhatsApp, and Twitter were the most used platforms.
The usage pattern of different social platforms by the participants is shown in~\cref{fig:platform_usage}.

\begin{table}[h!]
	\centering
    \caption{Participants' Demographics}
    \resizebox{0.85\columnwidth}{!}{
        \begin{tabular}{p{15em}cc}
        \toprule
                                                             &   \textbf{US} \# (\%) &  \textbf{Europe} \# (\%)\\
        \midrule
        \textbf{Gender}                                      &          &           \\
        \quad Female                                               & 48 (52\%)      & 49 (50\%) \\
        \quad Male                                                 & 42 (46\%)      & 49 (50\%) \\
        \quad Other                                                & 2  (2\%)       &  --- \\
        \hline
        \textbf{Age}                                      &          &           \\
        \quad 18 - 24                                           & 29 (31\%)     & 22 (22\%)    \\
        \quad 25 - 34                                           & 36 (39\%)     & 39 (40\%)   \\
        \quad 35 - 44                                           & 18 (19\%)     & 21 (21\%)   \\
        \quad 45 - 54                                           & 4 (4\%)       & 12 (12\%)   \\
        \quad 55 - 64                                           & 4 (4\%)       & 3 (3\%)    \\
        \quad 65 - 74                                           & 1 (1\%)       & 1 (1\%)   \\
        \hline
        \textbf{Education}                                   &          &           \\
        \quad Bachelor degree                                      & 35  (38\%)     & 34 (35\%)  \\
        \quad Some college---no degree                             & 20  (22\%)     & 22 (22\%)  \\
        \quad Graduate degree                                      & 15  (16\%)     & 20 (20\%)  \\
        \quad High school degree                                   & 11  (12\%)     & 15 (15\%)  \\
        \quad Associate degree                                     & 10  (11\%)     & 4 (4\%)   \\
        \quad Less than high school                                & 1   (1\%)      & 2 (2\%)   \\
        \quad Prefer not to answer                                 & ---            & 1 (1\%)   \\
        \hline
        \textbf{Marital Status}                              &          &           \\
        \quad Single, never married                                & 55  (60\%)     & 49 (50\%)  \\
        \quad Married/domestic partner                             & 31  (34\%)     & 44 (45\%)   \\
        \quad Divorced                                             & 4   (4\%)      & 4 (4\%)   \\
        \quad Separated                                            & 2   (2\%)      & ---       \\
        \quad Prefer not to answer                                 & ---            & 1 (1\%)   \\

        \hline
        \textbf{Income}                                      &          &           \\
        \quad \$0 - \$19,999                                          & 9 (10\%)    & 22 (22\%)    \\
        \quad \$20,000 - \$39,999                                    & 18 (20\%)    & 26 (26\%)   \\
        \quad \$40,000 - \$59,999                                    & 14 (15\%)    & 16 (16\%)   \\
        \quad \$60,000 - \$79,999                                    & 15 (16\%)    & 13 (13\%)   \\
        \quad \$80,000 - \$99,999                                    & 9 (10\%)     & 7 (7\%)    \\
        \quad \$100,000 or more                                      & 23 (25\%)    & 5 (5\%)   \\
        \quad Prefer not to answer                                   & 4 (4\%)      & 9 (9\%)    \\
        \hline
        \textbf{Background in IT}                            &          &           \\
        \quad Yes                                                  & 29 (32\%)      & 31 (31\%)   \\
        \quad No                                                   & 62 (67\%)      & 66 (67\%)   \\
        \quad Prefer not to answer                                 & 1  (1\%)       & 1 (1\%)  \\
        \bottomrule
        \end{tabular}
        \label{tab:demographics}
    }
\end{table}

\subsection{Analysis Method}\label{sec:analysis_method}

\paragraph{Coding Free Text Answers}
We coded free text answers obtained from our survey to uncover users' perceptions. In our analysis, two researchers independently coded free text responses using a shared codebook. Across questions, Cohen’s $\kappa$ (inter-rater agreement~\cite{landis-cook-agreement}) ranged from 0.7 to 1, indicating substantial to perfect agreement. The coders met to resolve disagreements and choose a final code.

\paragraph{Statistical Analysis}
We leveraged statistical hypothesis testing to investigate significant deletion privacy norms.
Specifically, for such analysis, we converted five-point Likert scales to the ordinal variable as follows: Completely Acceptable (2), Somewhat Acceptable (1), Neutral (0), Somewhat Unacceptable (-1), Completely Unacceptable (-2).
Unless otherwise stated, we used the nonparametric Mann Whitney U test to compare the 
responses across different groups. 
For all tests, the level of significance ($\alpha$) was 0.05 and further adjusted using Bonferroni multiple-testing correction.

\subsection{Ethical Considerations}

To adhere to the principles of ethical research we took the following steps.
In the recruitment process, each participant was informed
of the study's purpose, that they can withdraw at any time without giving any reasons, and that we would not store any personally identifying information.
We also informed the participants about the study's estimated duration
and their compensation in our consent form.
Respondents who did not consent were not allowed to proceed with the study.
Our study protocol was examined and approved by the lead author's Institutional Review Board (IRB).

\subsection{Limitations}
We did our best effort to plan and conduct the survey thoroughly; However, like all prior user studies, our results should be interpreted in context of its limitations. 

We used the Prolific Academic to recruit our participants, and obtained 191 participants which is in line with earlier applications of CI~\cite{apthorpe2019evaluating}. Our recruiting approach might have resulted in a younger and more tech-savvy sample which is not necessarily representative of the population. However, earlier work found that crowdsourcing for security and privacy survey results can be more representative of the US population than census-representative panels~\cite{redmiles-2019-mturkrepresent}. Further, responses from Prolific participants creates higher quality data than comparable platforms~\cite{peer2017beyond}.

As our survey and videos were in English, we required the participants to be fluent in English, which could have resulted in a language and consequently a cultural bias. However, our study sheds light on user perceptions about an under-studied privacy violation and we identified significant user concerns even within our sample. Thus, we strongly believe our study is useful to establish the importance of the problem and uncover normative rules governing deletion privacy. An in-depth understanding of the language or culture-specific variations of deletion privacy using participants from multiple languages/cultures is an intriguing future work for privacy researchers and complementary to this study.


Our survey videos were narrated by a non-native English-speaking researcher, which may have caused some issues with the accent and pronunciations.
We made the best effort possible by recording multiple times and narrating according to a script. Further, the videos have been uploaded to Youtube, where the participants could have used the subtitles if needed.

The survey was conducted in two parts (see~\cref{sec:survey_instrument}), with an average completion time of 25 minutes. The pilot study participants did not indicate any issue (due to participant fatigue) with the length of the survey (we added an explicit question in pilots). Still, to account for unforeseen participant fatigue, 
we made all the explanation questions (i.e., free text form questions) optional and only required the participants to respond to the multiple-choice and matrix questions (except for the usefulness question of the deletion mechanisms). In case the participant chose not to answer optional questions, the average completion time would have been less than 15 minutes. Furthermore, in the second part of the survey, we primarily leveraged videos (with subtitles) to explain the survey context and reduce the participant fatigue due to comprehending long descriptive text.

Part of this study aims to unpack the users' experiences with deletion events, which directed us to gather self-reported data from the participants. \newtext{Therefore, the effect of the \textit{after-the-fact} of the responses may have resulted in approximate answers to our questions.} Nevertheless, our results are in line with the prior studies.

\section{Results}
This section presents our findings on deletion privacy exploration.
We begin with the users' past deletion experiences.

\subsection{\hspace{-1mm}Users' Deletion Experiences-RQ1}\label{sec:result_history}
\begin{figure}[b]
	\centering
	\resizebox{0.85\columnwidth}{!}{
	    \includegraphics{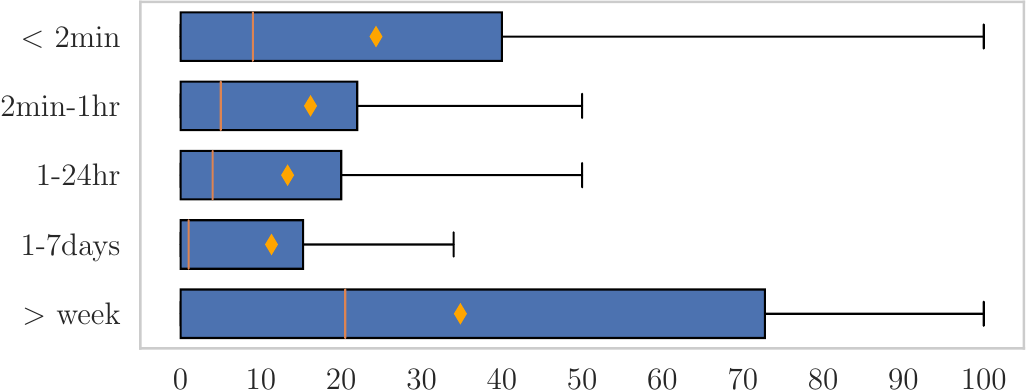}
	}
	\caption{{Boxplot of self-reported \% of deleted content, removed at different time periods after publishing the post. Diamonds indicate the average percentage for each period.}
	}
	\label{fig:deletion_frequency}
\end{figure}

\subsubsection{Many users delete their outdated posts}
Among the 191 participants, a significant majority of 82\% 
reported that they had deleted a post(s) in the past. 78\% report that they have deleted a post(s) from Facebook, 46\% from Instagram, and 34\% from Twitter. 
These numbers match with earlier work on social content deletion~\cite{tinati2017instacan,mondal-2016-longitudinal-exposure}.

We further asked the participants who have deleted posts in the past, to indicate how frequently they have deleted (all) their posts in different periods after publication (i.e., the percentages of deletions made in different periods).
The frequency results are shown in~\cref{fig:deletion_frequency}.

We see that 24\% of the deletions are within less than two minutes from their publishing time (similar to~\cite{tweets-forever}, that reports 22\% of deletions happening within a minute of appearing in Twitter), perhaps hinting at a spelling or grammar issue.

The ``2 minute - 1 hour'', ``1 - 24 hours'', and ``1 - 7 days'' periods have similar frequency percentages (11-16\%). These are periods where possible feedbacks are given by the users' family/close friend group (that closely follow the user's activities), coworkers/social friends (that check on their acquaintance's activities daily), and a much larger audience when posts go viral after a couple of days.   

Interestingly, the largest category belongs to the deletions ``after a week'' with more than one third (35\%) of all the deletions (similar to~\cite{mondal-2016-longitudinal-exposure}, where they report that one-third of all posts are deleted after six years of their publication).
This indicates that old posts on social platforms are not necessarily ignored, and users actively care about them and remove the unwanted ones.


\begin{table}[b]
    \small
	\caption{{Reasons for post deletions}}
	\vspace{-2mm}
		\centering

	\resizebox{0.85\columnwidth}{!}{
		\begin{tabular}{lc}
			\toprule
			\multirow{1}{*}{\textbf{Reasons}}    &  \# (\%) \\
			\hline
			{Being irrelevant due to time passing}          			&   100 (64\%) \\  
			{Fixing Spelling/Grammar/FactCheck}             		&   77  (49\%)  \\
			{Post did not get enough attention}             			 &   46  (29\%)    \\
			{Cleaning up my profile for new relationship}  	 		&   36  (23\%)    \\
			{Cleaning up my profile for new job}            			&   36  (23\%)    \\
			{Removing drug/alcohol/sexual related content}     &   18  (12\%)     \\
			{Removing Racial/Religious/Political content}    	  &   15  (10\%)  \\
			{Removing due to controversy/harassment}           &   13   (8\%)   \\
			{Removing violence/cursing related content }    	 &   11  (7\%)     \\
			{Removing embarrassing content}                			 &   10   (6\%)       \\
			{Removing health related content}              			   &   8   (5\%)           \\
			{Personal reason}                               					 &   7   (4\%)     \\
			{Other}                                         						    &   6   (4\%)  \\
			\hline
		\end{tabular}
	}
	\label{tab:deletion_reasons}
\end{table}

\subsubsection{Users delete their posts for non-obvious reasons}\label{sec:deletion_reasons}
We observed that over 80\% of the participants had deleted at least one of their posts within the social platforms. 
The next question that comes to mind is what are the reasons behind users' deletions.
We asked the 156 participants (that reported a deletion) about their reasonings. 
Nine different reasons (shown in \cref{tab:deletion_reasons}) were presented to them (taken from prior works~\cite{zhou-2016-tweetproperly,tweets-forever} and pilot studies) and further given an ``other'' option that they could have provided  additional reasons in a free text box. 
After categorizing the responses, we added four other reasons (i.e., ``removing due to controversy/harassment,'' ``removing embarrassing content,'' ``personal,'' and ``other'') to the list.

The participants reported that the main reason for deleting their posts was that they became irrelevant as time passed (64\% of the users). 
This reason highlights the fact that 35\% of deletions occurred after a week.
About half of the participants indicated that they had removed a post due to the obvious reason of fixing spelling/grammar and factual checking.
This reason is in line with the 24\% of deletions happening within a short time of publishing (less than 2 minutes).

In~\cref{tab:deletion_reasons}, we see that a significant number of the participants have reported sensitive topics  (drug, alcohol, race, politics, carnal, violence, etc.) as the reason for their deletion(s). 
Moreover, 23 of the participants (15\%) self-reported that they have deleted their posts to remove  contents that were embarrassing or caused some controversy and harassment.
 


		

\subsubsection{Do the content of the deleted posts contain sensitive or damaging information?}
We observed that users delete their posts for many reasons, and some can be considered sensitive to some people. Therefore, in the next question, we asked whether they agree or disagree with the following statement---``when someone deletes a social media post, it indicates that the content of that post is sensitive/damaging/embarrassing to that individual.'' Further, they were asked to provide an example in support of their answer.

More than half of the participants (i.e., 51\%), to some degree, agreed with the statement (i.e., strongly agreed or somewhat agreed). Among these participants, 74\% gave examples and details about a situation that their deleted post contained some embarrassing, inappropriate, or emotional content.  
For example, participant $P29$ wrote: \textit{``Someone would probably delete posts that would be embarrassing or legally damaging for the public to know about''}.

Conversely, 25\% disagreed (i.e., strongly disagreed or somewhat disagreed) with the statement, and 60\% of them gave examples about removing irrelevant content or fixing grammatical mistakes that contain no sensitive or damaging information.
E.g., participant $P61$ wrote: \textit{``I've posted things that looking back an hour later I just think are dumb, nothing embarrassing or offensive''}.

The remaining 24\% of the participants neither agreed nor disagreed. They indicated that it is dependent on the context of the posts and can be considered either damaging or non-damaging. Therefore, they chose to be neutral about the statement. E.g., participant $P3$ wrote: \textit{`Those could be the reasons why, but they could also just think it's irrelevant/outdated, stupid, not worth having up, or factually wrong.'}.

As we can see, users have different perspectives about the content of their deleted posts, but almost half of them consider the content to contain some damaging/sensitive information. Therefore, it is important that this information remains hidden and not be exposed to the public by some malicious entity that seeks the damaging deletions of the users.

\subsubsection{Who notices the users' deletions?}
Previously we saw that 82\% of the participants reported that they had deleted their posts in the past. 
To see if anyone has noticed these deletions, we asked those participants---``have you ever become aware of someone noticing that you have deleted one of your posts?''
The question was followed by asking them to identify social group(s) who noticed their deletion(s). The social groups were family members, friends, coworkers, and strangers---with varying closeness to the user.

Fifty-four participants (i.e., 35\%) self-reported that they are aware of situations where someone noticed their deletions.
The majority (50 out of 54, i.e., 93\%) of the participants' deletions were noticed by their friends' group. In second place with 41\%, (of 54 participants) family members were the ones that noticed the participants' deletions. Not surprisingly, only a small number of cases (9\%) were noticed by the strangers.

We repeated the two questions by changing the roles.
We asked the participants whether they had noticed anyone deleting their posts and who they were? 
Among 191 participants, 68\% (130) reported that they have noticed at least one other user's deletion(s).
Among the social groups that the participants noticed their deletions, the friend group stood out with highest percetage of 70\%. 
The second highest-ranked group was the ``stranger'' group (42\% of participants noticed deletions from strangers); only 17\% reported noticing their family members' deletions. Thus, a significant fraction of our participants notice strangers' post deletions, suggesting the possibility of deletion privacy violation for those post deleters. 

We note that in our study, we did not specifically ask the users to name the individuals that they have noticed their deletions for various privacy reasons. Therefore, we do not rule out the possibility that the users may have notice deletions from high profile users such as celebrities and politicians and as a result, reporting higher percentages in the stranger group.

\subsection{\hspace{-3mm}Contextual Norms of Deletion Privacy-RQ2}\label{sec:result_privacy}
Next, we analyze the contextual norms of deletion privacy using data collected from the CI-driven questions. 
The outcome (dependent) variable for CI questions was the users' acceptability score for each information flow, ranging from -2 (completely unacceptable) to 2 (completely acceptable) as explained in~\cref{sec:analysis_method}.
We used a multivariate regression model to compare and discuss the different CI parameter values and significance to the acceptability score (dependent variable). \cref{tab:regression_results} presents the highlighted model results.

\begin{table}[]
    \centering
    \caption{Regression model results for the CI parameters and demographics. In the table we only show the variables that have significant difference with the baseline variable values.}
    \resizebox{\columnwidth}{!}{
    \begin{tabular}{l c c c c}
         \toprule
         \textbf{Variable} & \textbf{Coefficient} &\textbf{ 2.5\%} & \textbf{97.5\%}  & \textbf{p-value}\\
         \toprule
         \textbf{Baseline:}\\
         \multicolumn{5}{l}{\textbf{Subject=\textit{null}, Recipient=family, Transmission=\textit{null}, Attribute=irrelevancy}}\\

         Subject = self             &   -0.306   &   -0.411  &   -0.202  & <0.001*** \\
        \hdashline
         Recipient = coworkers      &   -0.363 &   -0.478  &   -0.248  & <0.001*** \\
         Recipient = company        &   -0.737 &   -0.851  &   -0.624  & <0.001*** \\ 
         Recipient = government     &   -1.132 &   -1.245  &   -1.018  & <0.001*** \\
         \hdashline
         Transmission = interacted  &   0.313  &   0.232   &   0.394   & <0.001*** \\
         Transmission = observing   &   0.0981  &   0.016  &   0.18   &  0.019*  \\
         \hdashline
         Attribute = not enough attention   & -0.449 &-0.597&   -0.302  & <0.001*** \\
         Attribute = spelling/grammar& 0.2902 & 0.143&   0.437   & <0.001*** \\
        \midrule
        \midrule
        \textbf{Baseline:}\\
        \multicolumn{5}{p{\columnwidth}}{\textbf{Gender=Male, Age=35+, Education=university,   \newline Income=<\$70K, Marital=single/relation}}\\
        
        Gender = Female             &   -0.0418   &   -0.112   &   0.029   & 0.246 \\        
        Age = 18-35                 &   0.0677    &   -0.007   &   0.142    & 0.075 \\
        Education = university degree&  -0.0201   &   -0.092   &   0.051   &  0.582 \\
        Income = more than \$70K    &   -0.1419   & -0.216   &  -0.068  & <0.001*** \\
        Marital = separated/divorced&   -0.3609    &-0.519   &  -0.203  & <0.001*** \\
        \bottomrule
        \multicolumn{5}{r}{Significance codes: ***$p$< 0.001, **$p$<0.01, *$p$ < 0.05}
    \end{tabular}
    }
    \label{tab:regression_results}
\end{table}



	
    

\subsubsection{Deleted posts that are about the users themselves need more protection}
One of the CI parameters that we explored in this study is the ``subject'' of the deleted posts. 
The only ``subject'' that has a significant difference ($p$ < 0.001) with the baseline \textit{null} subject (i.e., not specifying any particular subject) is the ''user itself.'' We further see this point in the average score of the flows. On average, the flows with ``user itself'' as the subject had a 0.35 Likert-scale point less acceptability.  This point shows that users are more concerned about the deleted posts with themselves as subject than anyone else.

\subsubsection{Users seek deletion privacy against large-scale data collectors}
We found a statistically significant difference for the recipients of the deletion events when the users’ deletions are noticed by their outer social circles such as the government, a company, or even their coworkers compared to their friends and family members (\cref{tab:regression_results}).
The average acceptability score of information flows that have ``the government'' as their recipient is mostly negative, 
indicating that most of the participants consider these flows as ``completely unacceptable'' or ``somewhat unacceptable.'' Although less severe, the same is true for ``a company'' as the recipient. We observe in~\cref{tab:regression_results} that the coefficients of these two cases have the highest negative magnitude ($p$ < 0.001). 
Further, we observe that in the case of ``coworkers,'' there is also a significant difference with the baseline ``family,'' but its coefficient magnitude is smaller and affects the model outcome less significantly.

The average scores of flows with the recipient ``family,'' ``friends,'', and ``coworkers'' are all positive and mostly above one (except ``coworkers,'' which is between zero to one). 
On average, flows with large-scale data collectors as the recipient (a company or the government) are 0.94 Likert-scale points less acceptable than their other recipient counterparts.
This result provides quantitative evidence that social platform users are indeed concerned about their deletions being noticed by third-party services and state agencies; Therefore, there is a need  for the social platforms 
to proactively create different deletion privacy policies for different classes of recipients.


\subsubsection{Not knowing how deletions were noticed is less acceptable to the users}
In~\cref{tab:regression_results}, we see that both of the transmission principles that we considered in our study are significantly different ($p$ < 0.05) that the baseline \textit{null} transmission (i.e., not specifying any discovery method to the participants). On average, the score of the flows with the null transmission principle is 0.23 Likert-scale points smaller (less acceptable) than the non-null transmission principles. This is consistent with the human desire for cognitive closure~\cite{Congnative} that not knowing how deletions were noticed is less acceptable.

\subsubsection{Users want to hide their non-popular posts more than any other post}
In this study, the \textit{attribute} parameter of the CI flows corresponds to the reasons for removing a post. 
We identified ten different reasons for the deletions shown in~\cref{tab:ci_infoflow} (bottom portion). The baseline attribute considered in the regression model was ``Being irrelevant due to time passing,'' as it has been identified as the most common reason for the users to delete their posts (see \cref{sec:deletion_reasons}).
The two attributes that were significantly different ($p$ < 0.001) were ``Fixing Spelling/Grammar'' and ``Post did not get enough attention'' (\cref{tab:regression_results}). These two deletion reasons are at the opposite ends of the acceptability scores. Flows with the attribute ``Fixing Spelling/Grammar'' have the highest average acceptability score (0.98 on the Likert-scale points), and the flows with the attribute ``Post did not get enough attention'' have the lowest (0.24 on the Likert-scale points). 

We further analyzed the low-scored flows by checking the correlation between the scores given by users who had self-reported deleting content for that reason. Interestingly, for ``Post did not get enough attention,'' the negative scores primarily came from people who did not delete content because of that reason. This finding hints that our participants might have \textit{perceived} digging up forgotten posts by virtue of deletion as a 
violation of deletion privacy. In other words, the platforms should consider providing stronger deletion privacy to non-popular posts than popular posts.




\subsubsection{Effect of demographics on deletion privacy}\label{sec:demographics_results}
We examined the influence of common demographics (i.e., gender, age, education, income, and marital status). We present the effects and significance for each of the categories in~\cref{tab:regression_results}. We observed no significant difference between the male and female participants, their age, and their education levels. In the following, we discuss the details for the other demographic categories.

\paragraph{Higher-income users are more concerned about their deletions, but no difference for different education levels}
We asked about the users' household income in intervals of \$10,000. To have a meaningful analysis, we considered the US median household income of \$68,703~\cite{censusincome}, and as a result, set our splitting point at \$70,000.
We see a significant difference between the two groups ($p$ <0.001), where users with higher incomes are more worried about their deleted posts with lower average scores. 

We further divided the users based on their education degree, where one group had obtained a university degree (i.e., associate, undergraduate, and graduate degrees) and another which did not. However, we saw that there is no significant difference between the two groups.



\paragraph{Individuals that have ended their relationship in the past are more conservative}
We found that, on average, individuals that had identified their marital status as divorced or separated had a lower acceptability score (average of 0.22 Likert-scale points) compared to the individuals that identified themselves as single, never married, or in a current relationship (average of 0.62 Likert-scale points).

\subsubsection{Differences between the US and Europe}\label{sec:diff_us_europe}
We note that there is a great diversity of privacy expectations across geopolitical boundaries, and considering entire Europe as a monolith may not capture all the details. However, we strongly feel that tackling finer-grained group-specific (for multiple demographic/societal groups) norms of deletion privacy is a great future direction that is out of scope for this study.
Nevertheless, in what follows, we present the differences observed in the results of the CI-driven questions between the participants from the US and Europe.

\paragraph{Coworkers are considered a closer social group in the US}
We applied a statistical test between the flows of the recipients ``coworker'' and ``family'' for the US participants and saw no significant difference.
However, considering the same condition, there is a significant difference for the European participants ($p$ < 0.001). 

The opposite is true when comparing the distribution of the flows where the recipients are ``coworker'' and ``a company.''  
In this case, there is no difference between the distributions in Europe, but one exists for the US ($p$ < 0.001).
We conclude that individuals consider their coworkers to be in a closer social group in the US compared to European individuals.

\paragraph{Stalking is more of a concern in Europe}
Earlier, we saw that users are much more comfortable knowing what transmission principle (method of noticing the deletion) is used to notice their deletions. 
However, we observe a difference between the acceptability of the US users and European users when looking at the transmission principle ``checking and observing the user profile regularly'' (in other words stalking). We compared the scores that participants in the US and Europe gave to the two defined transmissions separately.
We observe no significant difference between the two transmission principles for the US participants.
However, there is a significant difference between checking/observing the users' profiles regularly (stalking) and interacting with the post in Europe ($p$ < 0.001). 
Thus stalking is seen as a less acceptable mean of noticing deletions compared to noticing due to a prior interaction with a difference of 0.4 Likert-scale points. 

\paragraph{Demographics affect differences}
Previously, overall we did not observe significant difference between scores of female and male participants. However, when we separate the US and Europe participants, we observe that in Europe  gender is correlated with acceptability scores ($p$ < 0.01). European female participants are more conservative about their deletions, and on average, have $0.2$ lower Likert-scale point score.  

We see a similar trend for the older participants in Europe, where they are more concerned about their posts being noticed by others.  
We observe a significant difference between the average scores of participants that identified themselves between the age of 18-34 to all other older participants.
In fact, the younger generation (millennials and generation Z) has a higher acceptability score with an average of 0.65 Likert-scale points compared to all other participants (average score 0.46). This finding is supported by earlier research on the data-sharing behavior of teens~\cite{teens-pew}. 

\subsection{Evaluating Deletion Mechanisms-RQ3}\label{sec:result_mechanisms}
So far, we have established the need for deletion privacy and uncovered its norms on social platforms. In this section, we will compare the utility of different deletion mechanisms for enhancing deletion privacy in the presence of a large-scale adversary. We begin by investigating if the participants have ever experienced a negative scenario with the explained adversary in~\cref{sec:adversarial_model}.

\subsubsection{Some users have been attacked}

As detailed in~\cref{sec:survey_instrument}, part II of the survey began by explaining the malicious entity (see~\cref{sec:adversarial_model}) considered in this work, followed by asking the participants whether they have had any negative experience (issues/problems/discomforts) facing such a malicious entity in any of their social platforms? If they gave a negative response, we asked them to rate the likelihood of this scenario happening to them. On the other hand, if they indicated that this attack has happened to them, we requested an optional brief explanation of the negative experience.
95\% of the participants responded with ``No''; 
However, 35\% (of this 95\%) think that this scenario is likely to happen to them. 34\% responded that they do not think it is a likely scenario, and the remaining 31\% were unsure.   

Unfortunately, many of the 5\% of participants who had negative experiences did not provide details to the incident to extract meaningful patterns (likely because providing free-form explanatory responses was not forced).
However, out of the responses we received, Participant $P70$ wrote---\textit{``I saw one of my deleted photos on a Pinterest account that was not mine. It seemed like it was some kind of ad for earrings but it made me a little uncomfortable.''}
Further, 
Participant $P36$ wrote about an experience that one of his/her friends had in encountering such an attacker 
---\textit{``... such malicious entities surely exist. My friend was contacted by one and threatened that the malicious entity would send pictures to his mother''}.

\subsubsection{Selective deletion mechanism used by many social platforms is ineffective in hiding the deletions}

\begin{figure}[t!]%
	\centering
	\resizebox{\columnwidth}{!}{\includegraphics{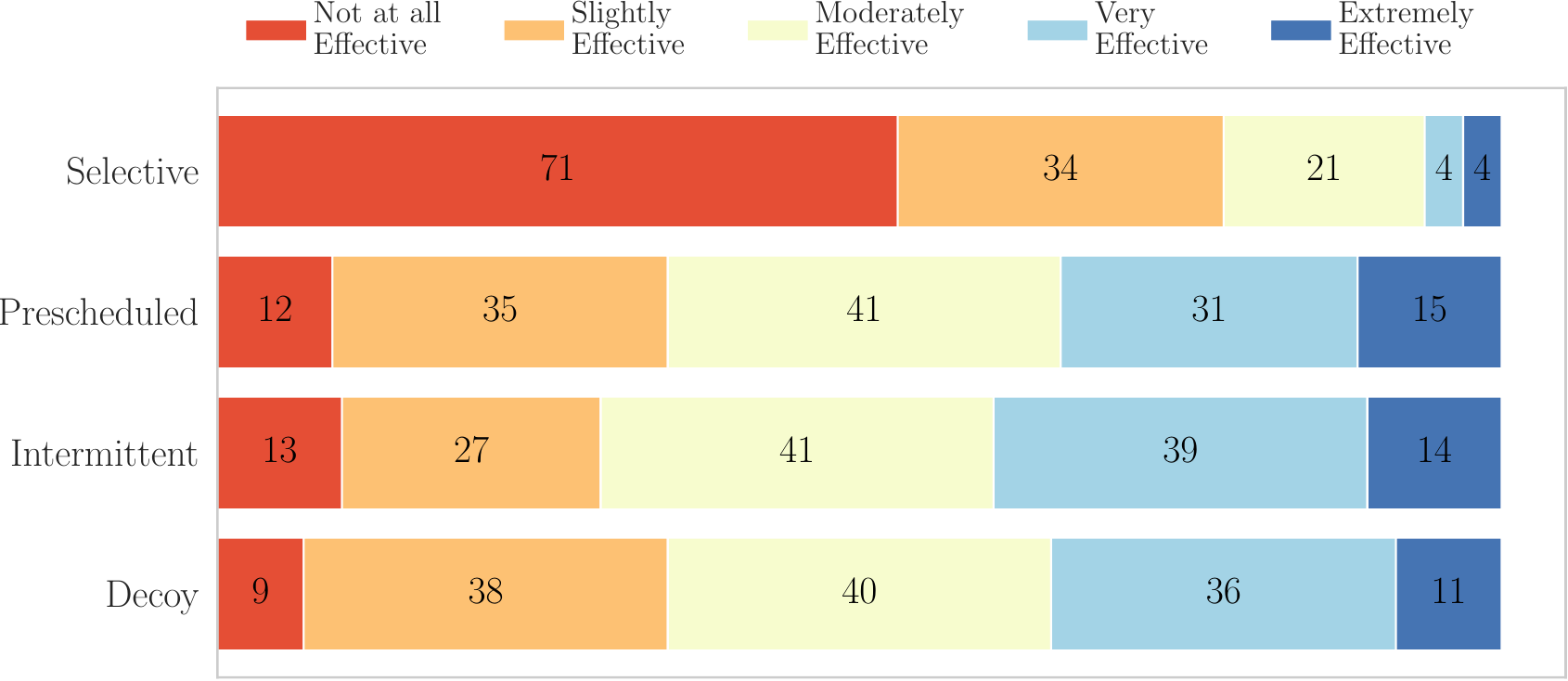}}
	\caption{{Effectiveness of deletion mechanisms in hiding the damaging/sensitive content of the users.}}
	\label{fig:sol_effective_all}
\end{figure}

For each of the deletion mechanisms (explained in~\cref{sec:mechanism_background}): Selective deletions, Prescheduled deletions, Intermittent Withdrawal, and Decoy deletions, a short video was shown to explain the mechanism and its characteristics. 
Next, we asked---``In your opinion, how effective is [Deletion Mechanism] in hiding your damaging/sensitive posts in the presence of a malicious entity who collects all deleted posts from a large number of users?''
We used a Likert scale for the responses: Not Effective at all (0), Slightly Effective (1), Moderately Effective (2), Very Effective (3), Extremely Effective (4).
The results are depicted in~\cref{fig:sol_effective_all}.

We observe that more than half of the participants indicated that Selective deletion is not effective at all for hiding the damaging deletions.
On the other hand, for Prescheduled deletion, only 9\% of the participants found it to be ineffective, but the remaining 91\% think it is somewhat effective.
Further, we statistically compared the effectiveness of the four different deletion mechanisms by applying the Wilcoxon signed-rank test to find the likelihood that these two groups of scores come from the same distribution. 
The results show that ``Selective deletion'' (i.e., used by many social platforms today) has a significantly different distribution from all the other three mechanisms 
with a mean of 0.77 Likert-scale points compared to 2.01.
The remaining deletion mechanisms do not have a significant difference among each other.

\begin{table*}[b!]
	\vspace{3mm}
	\caption{{Deletion mechanisms' characteristics.}}
	\vspace{-1mm}
	\resizebox{\textwidth}{!}{
		\begin{tabular}{l|cccc||cccc cc}
			\toprule
			\textbf{ Deletion Mechanism}  	& \thead{Maintain\\ Privacy} 	&  \thead{User in\\ Control} 			   &  \thead{Maintains\\ Archive} & \thead{No Assistance\\ Needed}   
			& \thead{No Privacy}			   &  \thead{Limited\\ User Control}  &  \thead{No Archive}      	  & \thead{Need of \\Assistance}    \\
			\midrule
			Selective Deletions~\cite{mondal-2016-longitudinal-exposure}				&   11 (8\%)						&				 \textbf{28 (21\%)  }		& 				7 (5\%)		  &        	---	   
			&   \textbf{95  (70\%)}		  &				 ---  	    				 		& 				---			    &        	---	  				   \\  
			
			Prescheduled Deletions~\cite{mondal2017longitudinal}     &   \textbf{63  (47\%)}			&				 ---				     		& 				---				   &              ---	   
			&   		18  (13\%)			   &				 14 (10\%)  			   & 	\textbf{56 (41\%)}		&              	---	   				 \\  
			
			Intermittent Withdrawals~\cite{minaei2019lethe} 	 &   \textbf{77  (57\%)}		 &				 4 (3\%)  					  & 			  7 (5\%)			&              	1 (1\%)	   
			&   23  (17\%)					   &		\textbf{29 (21\%)}  			& 				---					&              	--- 				  \\  
			
			Decoy Deletions~\cite{minaei2020deceptivendss}			  &   \textbf{66  (49\%)}		   &				 1 (1\%)  					& 				1 (1\%)			  &              	---   
			&   15  (11\%)		    		   &				 10 (7\%)  					 & 				9 (7\%)			 &        \textbf{27 (20\%)}    \\  											
			\bottomrule
		\end{tabular}
	}
	\label{tab:deletion_methods_useful}
\end{table*}
\subsubsection{Characteristics of the mechanisms useful to users}\label{sec:usefulness_all}
In addition to the effectiveness question, in the form of a free-text box, we asked the participant to describe cases where they find the mechanisms to be useful and/or NOT useful to them.
We categorized the responses into eight categories, where each corresponds to a characteristic of a mechanism in~\cref{tab:deletion_methods_useful} (An additional ``other'' category is omitted from the analysis).
The first four characteristics cover the positive aspects of the deletion mechanisms, and the second four characteristics point out their shortcomings. 

Previously, we observed that ``Prescheduled Deletions'', ``Intermittent Withdrawals'', and ``Decoy Deletions'' all have the same effectiveness in terms of protecting the users' damaging/sensitive deletions.
In fact, providing privacy to the removal of sensitive content was noted as the highlight of the mechanism in the usefulness question. 
However, as we see in~\cref{tab:deletion_methods_useful}, each mechanism has a particular deficiency that may become a barrier for their use.

\vspace{1em}
\paragraph{Prescheduled Deletions~\cite{mondal2017longitudinal}}
For this mechanism, participants particularly disliked the fact that the platform will not have an archive of their posts and eventually everything is deleted.
Participant $P19$ states: \textit{``It could be EFFECTIVE (and thus, useful) because it would take care of the issue of sensitive material being used maliciously. However, effective doesn't mean that I like the idea of it! It would NOT useful because I like the idea of having access to my old content (great for memories, etc.) and do NOT like the idea of losing it forever because of some system.''}

\paragraph{Intermittent Withdrawal~\cite{minaei2019lethe}}
In this mechanism, all the non-deleted posts are intermittently hidden for some amount of time by the system. Therefore,  the users felt a lack of control over their posts and profiles.
Participant $P15$ said: \textit{``This system could be useful if i made a sensitive post that I later decided to delete.  However, it could also be problematic if a social platform randomly made an important post that I needed my audience to see, invisible for a period of time.''} 

\paragraph{Decoy Deletions~\cite{minaei2020deceptivendss}}
In this mechanism, for each damaging/sensitive post, a set of decoy posts that are not damaging/sensitive to their owners (other users in the system) are selected to be deleted with the true damaging post. This procedure confuses the malicious entity in distinguishing which of the posts in the deleted set are the damaging/sensitive posts.
Although many users found this tactic effective and novel, the dependability on a pool of decoy posts from other users
prevented them from finding the mechanism useful. 
Participant $P119$ stated: \textit{``this technique could be very effective if you want to delete a post and protect it from the entity. However it is hard too find the decoy post.''}        

\paragraph{Selective Deletions~\cite{mondal-2016-longitudinal-exposure}}
As we observed previously, ``Selective Deletions'' was voted the least effective deletion mechanism in protecting the damaging deletions of the users. 
However, in the usefulness question, we see that it holds a unique characteristic that users admire.
Giving the users full control of their posts and profile seems to be an advantage of this mechanism.
Participant $P46$ stated: \textit{``It's not effective at all but it is however the most popular among the bunch listed. People (even me) like to have full control over our social medias and tweets. Regardless if a malicious bot tries to collect sensitive information off of us.''}         

To no surprise, we see that in some cases, users will sacrifice their privacy over the usability of the system. This work is an initial step towards discovering the needs of the users and maintaining a balance between usability and deletion privacy.

\section{Discussion and Future Work}\label{sec:discussion}
\oldtodo{some summary of the findings so far (if space)}

\subsection{Users Deeply Care about Their Old Posts}\label{sec:archival_importance}
In~\cref{sec:result_mechanisms}, we observed that the major hurdle for using ``Prescheduled Deletions'' is the lack of archival posts.
Before the deployment of the survey,  we suspected that it would be a concern for the users.
To that end, we added the following two questions to observe the importance of post archives. 

First, we asked---``How important is it for you that the social platform archives all your posts (new and old) and gives you the ability to access/view them at any time?''.
74\% of the participants stated that having access to their posts at a later time has some level of importance (24\% extremely important, 33\% very important, 17\% slightly important).
The remaining 26\% was split between 15\% neutral and 11\%, not at all important.

We further asked---``How important is it for you that the social platform archives all your posts (new and old) and gives others (i.e., those whom you have given permission to) the ability to access/view them at any time?''.
44\% of the participants stated this access pattern has some level of importance (9\% extremely important, 14\% very important, 21\% slightly important) to them.
The remaining 56\% was split between 26\% neutral and 30\%, not at all important.

This highlights the fact that users care about their old posts and want to be able to access their own posts and want others (to some level) to be able to access them at later times.

\subsection{A Significant Number of Users' are Willing to Help to Enhance Deletion Privacy in Decoy Deletion Mechanisms}
Earlier in this section, we saw that ``Decoy Deletions'' benefits from a pool of volunteer posts to provide privacy to the sensitive/damaging deletions. We further saw in~\cref{sec:usefulness_all} that users were worried about the need for assistance from other users' of the system for their sensitive/damaging deletions. 
Once again, before the survey's deployment, we suspected that the construction of the decoy pool could be a burden for some of the users.
However, to see the users' willingness to protect the damaging deletions of themselves and others, we asked the participants---``Imagine that Decoy Deletions is available to you on a platform. Would you be willing to offer some of your {non-sensitive/non-damaging posts that you won't mind getting removed from your profile} to be added to the decoy pool in order to protect the sensitive/damaging deletions of yourself and other users?''.
41\% of the participants responded Yes (12\% definitely yes, 29\% probably yes), 39\% of the 
participants responded No (13\% definitely no, 26\% probably no), and the remaining 20\% responded with ``might or might not''.

This shows that although the construction of the decoy pool and the need for assistance from other users is a concern for some users, there are a significant number of participants that are willing to contribute to the pool.

\subsection{Avenues to Improve  Deletion Mechanisms}\label{sec:future_work}
Before ending the survey, we wanted to capture whether participants could think of other solutions for protecting sensitive deletions on social platforms. 
To that end, we asked the participants---``Can you think of any other technique that can protect user deletions in the presence of the malicious entity?''

Many of the participants did not find any of the deletion mechanisms to be very effective 
(see~\cref{fig:sol_effective_all}), 
and 48.8\% (66 out of 135) of our participants proposed ideas to  address the problem of deletion privacy violation.
Out of them, 43.9\% (29 participants) suggested that the deletion problem can be erased altogether if users refrain from posting regrettable content and only maintain private accounts. Although this may be the most effective solution to the problem of deletion privacy, it is the most impractical one as users cannot accurately predict what content would be damaging to them in the future (e.g., before applying for a job or after a relationship breakup)~\cite{wang2011regretted,twitter13,mondal-2019-retrospective-facebook}.

\begin{table}[tb]
    \small
	\caption{{Users' suggestions for other deletion mechanisms}\vspace{-2mm}}
	\resizebox{\columnwidth}{!}{
		\begin{tabular}{l|cc}
			\toprule
			\textbf{ Deletion Mechanism}  	& \thead{\# of Votes} 	&   \thead{User intervention }   \\
			\midrule
			Decoy Deletions Variation                &   10      &   Yes          \\
			Prescheduled Deletions Variation         &   5       &   Yes          \\
		    Intermittent Withdrawals Variation      &   2       &   Yes          \\
			Proactive approaches			   		&   8 		&	Yes	         \\	
			Rate-limiting the adversary				&    8 		&	No	         \\
			Text Morphing				  			&    4 		&	No	  	     \\
			\bottomrule
		\end{tabular}
	}
	\label{tab:other_solutions}
	\vspace{-2mm}
\end{table}

\cref{tab:other_solutions} summarizes the remaining responses. 
We divide these user-proposed designs into two dimensions: i) variation of existing mechanisms, and ii) new proposed mechanisms.
We note that in-depth design, deployment, and evaluation of these mechanisms is beyond this work. Instead, these proposals paves way to promising future work and concrete actionable design considerations for future platform developers.

\subsubsection{Variation of existing mechanisms}
\vspace{-1mm}

\paragraph{Variation of Prescheduled Deletion}
Participants considered Prescheduled Deletion as one of the more effective solutions for hiding unwanted information. However, as we observed in~\cref{sec:archival_importance}, the importance of having an archive of the old posts outweighs the privacy implications. 
Therefore, to no surprise, participants suggested improvements that allows for some selective archival procedure for at least the post owner. 
Participant $P16$ goes one step further and requests the post's availability for those who have interacted with the post as well---\textit{``After a certain period of time/activity, all public posts automatically turn into private posts that can only be viewed by the owner and those who interacted with it.''}.

\textit{\textbf{Future design implications:}} Instagram recently started to provide similar functionalities and allows users to archive their stories that can be accessed later. However, evaluating the usability of this mechanism is a promising future work.

\vspace{3mm}
\paragraph{Variation of Intermittent Withdrawal}
Participants felt that the main downside of intermittent withdrawal is that the non-deleted posts are affected (i.e., they are hidden periodically), and users have limited control over them. 
To that end, participant $P24$ suggested that the intermittent withdrawal process should happen only when others are not viewing the user's profile (e.g., during late at night). 

\textit{\textbf{Future design implications:}} The user proposal (restricting the time of hiding)  helps utility and usability of the mechanism but  significantly affects deletion privacy--once the deterministic intermittent withdrawal period ends (i.e., in the morning), the deleted posts will be immediately revealed.

%
A more effective improvement 
for future designers might be to delay the intermittent process for the newly-created posts for a cool-down period, since the majority of a posts' interactions and views happen within the first few hours or days after post upload. 
However, there is a trade-off---if within this initial cool-down period, users decide to delete their posts their deletion privacy will be violated. Identifying this trade-off between right duration of cool down period and enabling the protection of intermittent withdrawal necessitates a system design consideration for the future designers. 

\vspace{3mm}
\paragraph{Variation of Decoy Deletions}
Decoy Deletions was one of the most interesting mechanisms for the participants, resulting in multiple proposed variants to improves usability. One prominent proposal was to reduce the need for assistance from other users by generating the decoy posts (e.g., random and non-sensitive posts or even contradictory posts) by posters themselves and then, at a later time, delete them altogether with the intended post. 


\textit{\textbf{Future design implications:}}
An interesting future work for resolving the problem of gathering decoy posts can be to use bot accounts to generate and disseminate synthetic decoy posts using generative language models like GPT-3 and BERT~\cite{devlin2018bert,brown2020language}. However, the key challenges---creating user-alike posts as well as the evading detection of bot accounts are both active research fields~\cite{stringhini2010detecting,chu2012detecting,dickerson2014using,wang2010detecting,bessi2016social,ferrara2016rise}.

\subsubsection{Newly proposed mechanisms}
\vspace{-1mm}
\paragraph{Proactive approaches to prevent the publication of sensitive content}
Eight participants pointed out different proactive approaches, similar to~\cite{wang2019donttweetthis, zhou-2016-tweetproperly}. In these proposals, multiple classifiers (e.g., Neural Networks, Naive Bayes, etc.) detect potential regrettable posts and advise users not to publish the posts. 
Participant $P186$ clearly explains the proactive solution in his/her response---\textit{``Some kind of bot/AI that, based on language and keywords, warn the user that their post may be considered offensive or sensitive before they post it in the first place meaning they can delete it before posting.''}.  

\textit{\textbf{Future design implications:}} Although helpful, in some cases, this proactive approach cannot prevent users from publishing future-regrettable posts~\cite{wang2011regretted}.
Furthermore, these approaches create overhead for the platform and slightly affect the freedom of publishing posts for users.

\vspace{2mm}
\paragraph{Rate-limiting the malicious entities}
Eight participants pointed out that the platforms should create barriers for the malicious entities that collect the users' data on a large-scale.
For example, participant $P108$ states---\textit{``Social networks could make efforts to thwart bots and scrapers that collect posts and also monitor profiles for deletions. Maybe IP limiting or some sort of CAPTCHA style tech?''}.

\textit{\textbf{Future design implications:}} In this approach, the adversary will not observe all the users' profiles constantly, or it will have blackout periods of the profiles (detecting deletions with significant delay).
However, this introduces a couple of challenges, such as the trade-off between transparency/openness and privacy. 
Furthermore, rate-limiting large-scale crawlers
in social platforms remains to be a challenge~\cite{vastel2020fp,mondal2012defending}.

\vspace{3mm}
\paragraph{Morphing the posts' text}
Finally, four of the participants suggest that users can edit their sensitive posts rather than deleting them. When they become comfortable enough with the edits, then they can either delete them or leave them on their profile. 
Participant $P62$ wrote---\textit{``Altering the post completely and then delete it. If tried to recover, the post would be completely different.''}

\textit{\textbf{Future design implications:}} This interesting proposal can be considered a feature of the platform itself, where this transition of the texts is automated. 
This process would involve syntactic (e.g., passivization, clefting, adjunct movement, etc.) and semantic (e.g., grafting, pruning, substitution, etc.) transformations.
If the text morphing is performed without the users' input, then the platform itself will not know what is sensitive to the users, which also removes the platform's trust.
The only burden that the users will face is that the audience of the posts at a later time will observe the morphed version of the post and not the exact original text of the user.
\section{Conclusion}
\newtext{In this study, we observed that the majority of users are deleting their posts.} 
There is a strong user-need for ensuring deletion privacy in social platforms, 
as users consider deletions a tool for removing sensitive, damaging, and embarrassing content.
Further, using contextual integrity, we demonstrated the context-dependency of the rules for preserving deletion privacy. 
The study identified that it is acceptable for the users if the one-hop individuals (family members, friends) on their social graphs become aware of their deletions but not the large-scale data collecting actors (e.g., web-service data collectors or the government).
Furthermore, we showed that ``Selective Deletions'' (the current deletion mechanism offered by many social platforms) are inefficient in protecting to the users' deletions. 
Finally, we highlighted the key factors that future social platform designers should consider for attracting the users while providing them deletion privacy.

\section*{Acknowledgment}
We thank our shepherd Apu Kapadia and the anonymous reviewers for their insightful comments and suggestions towards improving the paper quality. This work has been partially supported by the National Science Foundation under grant CNS1719196.

{
	\bibliographystyle{plain}
	\bibliography{main}
}

\appendix
\newpage

\newpage

\onecolumn

\section{Recruitment Message and Criteria Posted on Prolific Academic for Our Study}
\subsection{Recruitment Message}\label{sec:recruitment_message}
In this study, you will be asked a series of questions about your personal social media usage and the post deletions that you may have had or seen others make. To be eligible for this study, you need to be currently using one or more social platforms (e.g., Facebook, Twitter, Instagram, Whatsapp, Pinterest, Telegram, Slack, etc.). 

\noindent This survey is in the context of personal activities on social media platforms and NOT professional activities (e.g., managing your company's social media account).\\

\vspace{-2mm}
\noindent Participants need to be fluent in English to take part in this survey.\\

 \vspace{-2mm}
\noindent SURVEY CONTAINS SOME VIDEOS AND PARTICIPANTS NEED TO BE ABLE TO HAVE AUDIO CAPABILITIES TO LISTEN TO THE VIDEOS.\\

\vspace{-1em}
\subsection{Recruitment Criteria}\label{sec:recruitment_criteria}

\noindent We configured the survey on Prolific Academic platform (using Prolific Academic provided settings) such that our survey will only be advertised to participants who are: i) 18 years old or above, ii) had not taken our pilot-study, iii) had taken a minimum of 50 prior surveys on the platform, iv) had a minimum approval rate of 95\%, and  v) fluent in English. Note that, we did not deliberately recruited users who deleted their social media posts.\\

\vspace{-1em}
\section{Survey Instrument}\label{sec:survey_full}
In this section, we are providing the questions included in part 1 and part 2 of our survey. Part 1 of our survey contained two broader parts: Experiences about prior post deletions, and CI-parameter based questionnaire about deletion privacy. In the second part of the survey, we primarily presented four deletion mechanisms. Participants evaluated the usefulness of these mechanisms to preserve deletion privacy. We also collected participant demographics at the end of the Survey. 

In this section, we are also putting the branching logic and participant-question assignment logic used in this survey in  \textcolor{gray}{Gray} colored text.

\subsection{Part 1 of the Survey}


\hrule
\vspace{5mm}

\paragraph{Social Media Usage \& Post Deletions}\\
In this section we will ask you questions about your social media usage and whether you have ever deleted any of your social media posts. 
Throughout the survey when we refer to post deletions, we are referring to posts that have selectively been deleted and not those that were automatically deleted by the platform (e.g., Instagram/Facebook stories, Snapchat).
At the end of this section, we will also ask if you have ever noticed some other social media user deleting their posts and possible reasons behind the deletion.

\begin{enumerate}[start=1,label={Q\arabic*:}, leftmargin=*]
    \item \noindent Do you currently use any social media platform (e.g., Facebook, Twitter, Snapchat, Youtube, etc.), either via Mobile or via Web interface?  $\bigcirc$Yes  $\bigcirc$No\\
    \snote{If No is selected then skip to the end of survey.}\\
    
    \item \noindent What is the frequency at which you currently post (or in some platforms leave comments) in each of the following social media platforms? (mark "never used" if you have never used a platform)
   		\begin{compactitem}
   			\item (Matrix-style grid with the following rows): YouTube, Facebook, Instagram, Pinterest, Snapchat, LinkdIn, Twitter, WhatsApp, Slack, Telegram, Cybersocial, Other1[text box], Other2[text box], Other3[text box]
   			
   			\item Answer choices (columns) for each row: $\bigcirc$daily $\bigcirc$weekly  $\bigcirc$monthly $\bigcirc$yearly $\bigcirc$don't post/comment anymore $\bigcirc$deleted account $\bigcirc$never used
   		\end{compactitem}
   
   \item \noindent Have you ever deleted one of your social media posts in any of the previously mentioned platforms? $\bigcirc$Yes  $\bigcirc$No $\bigcirc$ I don't know\\
\end{enumerate}

\vspace{5mm}
\paragraph{Have Deleted Posts on Social Platforms as Mentioned in Q3 Above}\\
\snote{If the answer to the question 3 is Yes, we ask the following questions otherwise we skip to the next section.}
	
	\begin{enumerate}[start=4,label={Q\arabic*:}, leftmargin=*]
		
	\item \noindent Which platforms did you delete your social media posts from? (you may choose multiple platforms)\\
	$\square$ YouTube  $\square$ Facebook  $\square$ Instagram  $\square$ Pinterest  $\square$ Snapchat  $\square$ LinkdIn  $\square$ Twitter   $\square$ WhatsApp  $\square$ Slack   $\square$ Telegram   $\square$ Cybersocial   $\square$ Other1[text box]   $\square$ Other2[text box]   $\square$ Other3[text box]
    
    \item \noindent What were the reasons behind your deletion (pick as many as needed)?
    $\square$ Spelling/Grammar issue 
    $\square$ Cleaning up my profile for new job, 
    $\square$ Cleaning up my profile for new relationship
    $\square$ Being irrelevant due to time passing 
    $\square$ Racial/Religious/Political reason 
    $\square$ Removing sexual content
    $\square$ Removing drug/alcohol related content
    $\square$ Removing violence/cursing related content
    $\square$ Removing health related content
    $\square$ Post did not get enough attention
    $\square$ Prefer not to answer
    $\square$ Others (separate with comma)[text box]
    
    \item \noindent Please describe one such possible scenario where you have deleted a post. (in 1-3 sentences)\\
    
    \item \noindent Please indicate below how frequently you have deleted posts after less than 2 minute of posting, within 1 minute to 1 hour of posting, within 1 hour to 24 hours of posting and etc.
    For each of the cases indicate the approximate percentage using the sliders and the total should add to 100.\\
    \noindent [for each choice a slider: 0 . . . 100]
    
    \begin{compactitem}
    	\item deleted within less than 2 minute,  deleted within 2 minute - 1 hour, deleted within 1 - 24 hours,  deleted within  1 - 7 days,  deleted after a week
    \end{compactitem}

    \item \noindent Have you ever become aware of someone noticing that you have deleted one of you posts? $\bigcirc$Yes  $\bigcirc$No $\bigcirc$ I don't know
    
        \noindent \snote{If Yes is selected for question 8, we ask the following questions.}
        \begin{enumerate}[start=1,label={Q8.\arabic*:}, leftmargin=*]
            \item Below we put a few categories of people from your social circle. Please point out the people from each of the categories that have noticed your deletions (i.e., you were informed that they know about your deletion)? 
            You can select multiple categories. 
            $\square$ Family member
            $\square$ Friends
            $\square$ Coworkers/Acquaintances 
            $\square$ Stranger
           	$\square$ Prefer not to answer
            $\square$ Others[text box]
            
            \item Please describe one such scenario where some people from your social circle noticed your post deletion?
            (in 1-3 sentences)
            
            \item Did you face any issues/problems/discomforts due to others noticing your deletions?
            $\bigcirc$Yes  $\bigcirc$No $\bigcirc$ Prefer not to answer
 			
 			\snote{ If Yes is selected for question 8.3, then we ask:}
            \begin{enumerate}[start=1,label={Q8.3.\arabic*:},leftmargin=*]
                \item What issues/problems/discomforts did you face due to your post deletion? (Explain in 1-3 sentences)
            \end{enumerate}
        
            \snote {If Yes is NOT selected for question 8.3, then we ask:}
            \begin{enumerate}[start=1,label={Q8.3.\arabic*:},leftmargin=*]
                \item Suppose you were to delete one or more of your social media posts. What possible issues/problems/discomforts do you think you might face if you become aware of someone noticing your post deletion(s)?  (Explain in 1-3 sentences)
            \end{enumerate}
        \end{enumerate}

        \noindent\snote{If Yes is NOT selected for question number 8, then we ask the following questions.}
        \begin{enumerate}[start=1,label={Q8.\arabic*:},leftmargin=*]
            \item Below we put a few categories of people from your social circle. If you were to delete any of your social media posts, please point our how likely is it that people from each of the categories will notice your deletions?
            \begin{compactitem}
            	\item (Matrix-style grid with the following rows): Family member, Friends, Coworkers/Acquaintances, Stranger, Others[text box]
            	
            	\item Answer choices (columns) for each row:
            	$\bigcirc$Extremely unlikely $\bigcirc$unlikely $\bigcirc$Neutral $\bigcirc$likely $\bigcirc$Extremely likely
            \end{compactitem}

            \item Suppose you were to delete one or more of your social media posts. What possible issues/problems/discomforts do you think you might face if you become aware of someone noticing your post deletion(s)?  (Explain in 1-3 sentences)\\

        \end{enumerate}
            
\end{enumerate}

\newpage

\paragraph{Have NOT Deleted Posts on Social Platforms as Mentioned in Q3 Earlier}\\
\noindent\snote{If the answer to the question number 3 is NOT Yes, then we ask the following questions.}

\begin{enumerate}[start=4,label={Q\arabic*:}, leftmargin=*]
    \item If you were to delete any of your posts in any social media platform which platform would it be? (you can select multiple platforms)
    	$\square$ YouTube  $\square$ Facebook  $\square$ Instagram  $\square$ Pinterest  $\square$ Snapchat  $\square$ LinkdIn  $\square$ Twitter   $\square$ WhatsApp  $\square$ Slack   $\square$ Telegram   $\square$ Cybersocial   $\square$ Other1[text box]   $\square$ Other2[text box]   $\square$ Other3[text box]
    	
    \item If you were to delete any of your posts in any social media platform, what do you think the reason behind your deletion would be?  
    (pick as many as needed)
    $\square$ Spelling/Grammar issue 
    $\square$ Cleaning up my profile for new job, 
    $\square$ Cleaning up my profile for new relationship
    $\square$ Being irrelevant due to time passing 
    $\square$ Racial/Religious/Political reason 
    $\square$ Removing sexual content
    $\square$ Removing drug/alcohol related content
    $\square$ Removing violence/cursing related content
    $\square$ Removing health related content
    $\square$ Post did not get enough attention
    $\square$ Prefer not to answer
    $\square$ Others (separate with comma)[text box]
    
    \item Please describe one such possible scenario where you might delete a post? (in 1-3 sentences)
    
    \item Below we put a few categories of people from your social circle. If you were to delete any of your social media posts, please point out how likely is it that people from each of the categories will notice your deletions?    
	\begin{compactitem}
		\item (Matrix-style grid with the following rows): Family member, Friends, Coworkers/Acquaintances, Stranger, Others[text box]
		
		\item Answer choices (columns) for each row: 
		$\bigcirc$Extremely unlikely $\bigcirc$unlikely $\bigcirc$Neutral $\bigcirc$likely $\bigcirc$Extremely likely
	\end{compactitem}

    \item Suppose you were to delete one or more of your social media posts. What possible issues/problems/discomforts do you think you might face if you become aware of someone noticing your post deletion(s)? 
    (Explain in 1-3 sentences)
\end{enumerate}

\vspace{1mm}
\paragraph{Noticing Deletions Done by Others}

\begin{enumerate}[start=9,label={Q\arabic*:}, leftmargin=*]
    \item Have you ever noticed anyone deleting their posts?
   	$\bigcirc$Yes  $\bigcirc$No $\bigcirc$ Prefer not to answer
    
    \noindent \snote{If Yes is selected for Q9, we ask the following questions.}
    \begin{enumerate}[start=1,label={Q9.\arabic*:}, leftmargin=*]
        \item The individual(s) that deleted its post belongs to which of your social groups mentioned below? (you can choose multiple answers)
	        $\square$ Family member
	        $\square$ Friends
	        $\square$ Coworkers/Acquaintances 
	        $\square$ Stranger
	        $\square$ Prefer not to answer
	        $\square$ Others[text box]
        
		\item How did you become aware of the deletion(s)?
		 	$\square$ You were mentioned in the users' posts (e.g., tagged in a photo or your account ID was in the post )
			$\square$ Someone sent/mentioned the post to you
			$\square$ You interacted with the post by liking, commenting, reposting, sharing, etc.
			$\square$ You check the users' profiles regularly
			$\square$ Others (text box) 
		
        \item Please give an example in support of your answer to the above question. (in 1-3 sentences)
        
        \item Were you the subject of a deleted post made by another user (e.g., your name was mentioned in the post) or had an activity (liked, comment, repost, share, etc.) around a deleted post? $\square$ Yes $\square$ No
        
    \end{enumerate}
    
    \vspace{2mm}
    \noindent \snote{If Yes is NOT selected for  Q9, then we ask the following question.}
    \begin{enumerate}[start=1,label={Q9.\arabic*:},leftmargin=*]
        \item Below we are again putting a few categories of people from your social circle. What is the likelihood of you noticing any post deletion done by people from each of the categories?
        \begin{compactitem}
        	\item (Matrix-style grid with the following rows): Family member, Friends, Coworkers/Acquaintances, Stranger, Others[text box]
        	
        	\item Answer choices (columns) for each row:  $\bigcirc$Extremely unlikely $\bigcirc$unlikely $\bigcirc$Neutral $\bigcirc$likely $\bigcirc$Extremely likely
        \end{compactitem}
    \end{enumerate}
\end{enumerate}

\vspace{1mm}
\paragraph{Do Deletions Signify Damaging Posts}

\begin{enumerate}[start=10,label={Q\arabic*:}, leftmargin=25pt]    
    \item Do you agree or disagree to the following statement: when someone deletes a social media post, it indicates that the content of that post is sensitive/damaging/embarrassing to that individual.\\
    $\bigcirc$Strongly agree $\bigcirc$Somewhat agree $\bigcirc$Neither agree nor disagree $\bigcirc$Somewhat disagree $\bigcirc$Strongly disagree
    
    \item Please give an example in support of your answer to the above question. (in 1-3 sentences)
\end{enumerate}

\vspace{5mm}
\paragraph{CI-parameter based questionnaire about deletion privacy}

\noindent\snote{For this section, each participant was uniformly randomly assigned to a value for the \textit{SUBJECT} and \textit{TRANSMISSION PRINCIPLE} variable. That participant was also randomly assigned to a set of three \textit{recipients} variable values (out of two pre-defined sets). So for each each participant we can construct 3 possible combinations of <SUBJECT, TRANSMISSION PRINCIPLE, RECIPIENT> tuples. The set of all values for all of these variables are shown below.\\\\
	(Comma separated) list of \textit{SUBJECT} variable values: [that contained some information about yourself, that contained some information about your family members, that contained some information about your friends, that contained some information about your coworkers, \textit{null}]\\
	(Comma separated) list	\textit{TRANSMISSION PRINCIPLE} variable values: [because they were checking/observing your user profile regularly (to collect the public posts of users' and observing the removed posts), because they were mentioned in the post or interacted with the post (e.g., liking, commenting, reposting, sharing, etc.), \textit{null}]\\
	First set of \textit{RECIPIENT} variable values or \textit{Recipient\_A}: [your family member, your close friend, your coworker]\\ 	
	Second set of \textit{RECIPIENT} variable values or \textit{Recipient\_B}:[anyone, a company, the government]\\\\
The questions presented below was repeated three times by replacing the \textit{RECIPIENT} variable with the values from the assigned recipient set (Recipient\_A or Recipient\_B). However, the \textit{SUBJECT} and \textit{TRANSMISSION PRINCIPLE} remains same each time.
}\\

\noindent In the following we will show you 3 questions about the acceptability of others noticing your deletions under different scenarios. 
Note that the questions will all be similar and only differ on who will be the observer of the deletion.\\

\begin{enumerate}[start=12,label={Q\arabic*:}, leftmargin=25pt]    
	\item We are putting a few possible reasons behind post deletions in the table below (leftmost column). For each of these reasons, imagine a situation where you deleted a post [\textit{SUBJECT}] from one of your social media accounts due to that reason. 
	In each of these situations, please indicate how acceptable is it for you that [\textit{RECIPIENT}] notices your deletion [\textit{TRANSMISSION PRINCIPLE}]?
	(If you can think of other reasons please enter them in the “other” labeled text boxes and answer this same question)

	\begin{compactitem}
		\item (Matrix-style grid with the following rows, each for one value of the \textit{ATTRIBUTE} variable): 
		(i) Fixing Spelling/Grammar, 
		(ii) Cleaning up my profile for new job, 
	    (iii) Cleaning up my profile for new relationship,
		(iv) Being irrelevant due to time passing,
		(v) Racial/Religious/Political reason, 
		(vi) Removing sexual content,
		(vii) Removing drug/alcohol related content,
		(viii) Removing violence/cursing related content,
		(ix) Removing health related content,
		(x) Post did not get enough attention,
		(xi) Other1 [text box],
		(xii) Other2 [text box],
		(xiii) Other3 [text box]
		
		\item Answer choices (columns) for each row: $\bigcirc$Completely unacceptable $\bigcirc$Somewhat unacceptable $\bigcirc$Neutral $\bigcirc$Somewhat acceptable $\bigcirc$Completely acceptable
		
	\end{compactitem}
\end{enumerate}

\vspace{5em}
\subsection{Part 2 of the Survey}

\hrule
\vspace{5mm}
\paragraph{Describing Malicious Entity}

\noindent In this section of the survey, we describe different techniques that you may use to remove your posts from social media platforms and ask questions about the usage of those techniques.

For each of the techniques mentioned in this sections, consider the scenario that a malicious entity (not affiliated with the platform or users) is collecting the deleted posts of users from a social media platform that you use. 
The malicious entity's goal is to find some deleted posts that may be damaging/sensitive to the users and use them to harass or blackmail them. However, it does not have any background information on particular users (e.g., it does not know if a particular user will consider a post stating “I am smoking weed” or “I cannot trust anyone” as sensitive).

\noindent PLEASE WATCH ALL THE FOLLOWING VIDEOS COMPLETELY.

\begin{enumerate}[start=1,label={Q\arabic*:}, leftmargin=*]
	\item Please watch a short video about the malicious entity that we will consider for the rest of the survey.\\
	\url{https://youtu.be/isWxK1KTxN4}\\
	Have you ever experienced a scenario where a malicious entity who collects all deleted posts from large number of users caused any issues/problems/discomforts for you in any of the social media platforms?
	$\bigcirc$Yes  $\bigcirc$No
	
	\noindent \snote{If Yes is selected for Q1, then we asked the following question.}
	\begin{enumerate}[start=1,label={Q1.\arabic*:},leftmargin=*]
		\item What issues/problems/discomforts did you face? (Explain in 1-3 sentences)
	\end{enumerate}
	
	\noindent \snote{If No is selected for Q1, then we asked the following question.}
	\begin{enumerate}[start=1,label={Q1.\arabic*:}, leftmargin=*]
		\item Do you think that such a scenario can happen to you?  
			$\bigcirc$ Definitely yes
			$\bigcirc$ Probably yes
			$\bigcirc$ Might or might not
			$\bigcirc$ Probably no
			$\bigcirc$ Definitely no
	\end{enumerate}
\end{enumerate}

\vspace{5mm}
\paragraph{Comparison of Deletion Mechanisms}
\noindent\snote{The order of presenting the question blocks related to four deletion mechanism were randomized.}

\begin{enumerate}[start=2,label={Q\arabic*:}, leftmargin=*]
    \item Please watch a short video about a deletion technique called ``Selective Deletions''.\\
    \url{https://youtu.be/v-UxjiHhKhU}\\
    In your opinion, how effective is ``Selective Deletions'' in hiding your damaging/sensitive posts in the presence of a malicious entity who collects all deleted posts from a large number of users?
    $\bigcirc$ Not Effective at all
    $\bigcirc$ Slightly Effective
    $\bigcirc$ Moderately Effective
    $\bigcirc$ Very Effective
    $\bigcirc$ Extremely Effective

    \begin{enumerate}[start=1,label={Q2.\arabic*:}, leftmargin=*]
        \item Describe cases where the ``Selective Deletion'' technique might be useful and/or NOT useful to you.
        (explain in 1-3 sentences)
    \end{enumerate}    

    \item Please watch a short video about a deletion technique called ``Prescheduled Deletions''.\\
    \url{https://youtu.be/TW-4Tugtg9c}\\
    In your opinion, how effective is ``Prescheduled Deletions'' in hiding your damaging/sensitive posts in the presence of a malicious entity who collects all deleted posts from a large number of users?
    $\bigcirc$ Not Effective at all
    $\bigcirc$ Slightly Effective
    $\bigcirc$ Moderately Effective
    $\bigcirc$ Very Effective
    $\bigcirc$ Extremely Effective

    \begin{enumerate}[start=1,label={Q3.\arabic*:}, leftmargin=*]
        \item Describe cases where the ``Prescheduled Deletions'' technique might be useful and/or NOT useful to you.
        (explain in 1-3 sentences)

        \item How important is it for you that the social platform archives all your posts (new and old) and gives you the ability to access/view them at any time?
    	    $\bigcirc$ Extremely important
    	    $\bigcirc$ Very important
    	    $\bigcirc$ Neutral
    	    $\bigcirc$ Slightly important
    	    $\bigcirc$ Not at all important
    
    	\item How important is it for you that the social platform archives all your posts (new and old) and gives others (i.e., those who you have given permission to) the ability to access/view them at any time?
    		$\bigcirc$ Extremely important
    		$\bigcirc$ Very important
    		$\bigcirc$ Neutral
    		$\bigcirc$ Slightly important
    		$\bigcirc$ Not at all important
    \end{enumerate}    
    
    \item Please watch a short video about a deletion technique called ``Intermittent Withdrawals''.\\
    \url{https://youtu.be/nomlb8TEy9s}\\
    In your opinion, how effective is ``Intermittent Withdrawals'' in hiding your damaging/sensitive posts in the presence of a malicious entity who collects all deleted posts from a large number of users?
    $\bigcirc$ Not Effective at all
    $\bigcirc$ Slightly Effective
    $\bigcirc$ Moderately Effective
    $\bigcirc$ Very Effective
    $\bigcirc$ Extremely Effective
    
    \begin{enumerate}[start=1,label={Q4.\arabic*:}, leftmargin=*]
        \item Describe cases where the ``Intermittent Withdrawals'' technique might be useful and/or NOT useful to you.
        (explain in 1-3 sentences)
    \end{enumerate}
    
    \item Please watch a short video about a deletion technique called ``Decoy Deletions''.\\
    \url{https://youtu.be/HNCWlvtk1p4}\\    
    In your opinion, how effective is ``Decoy Deletions'' in hiding your damaging/sensitive posts in the presence of a malicious entity who collects all deleted posts from a large number of users?
    $\bigcirc$ Not Effective at all
    $\bigcirc$ Slightly Effective
    $\bigcirc$ Moderately Effective
    $\bigcirc$ Very Effective
    $\bigcirc$ Extremely Effective
    
    \begin{enumerate}[start=1,label={Q5.\arabic*:}, leftmargin=*]

        \item Describe cases where the ``Decoy Deletions'' technique might be useful and/or NOT useful to you.
        (explain in 1-3 sentences)

        \item Imagine that ``Decoy Deletions'' is available to you on a platform. Would you be willing to offer some of your non-sensitive/non-damaging posts that you won't mind getting removed from your profile to be added to the decoy pool in order to protect the sensitive/damaging deletions of yourself and other users?  
        $\bigcirc$ Definitely yes
        $\bigcirc$ Probably yes
        $\bigcirc$ Might or might not
        $\bigcirc$ Probably no
        $\bigcirc$ Definitely no

        \item Consider the case that the social platform (e.g., Twitter, Facebook) auto-creates posts for the decoy pool. In that case, would it be acceptable for you if the platform automatically publishes a random post on your profile, and at a later time (minutes to months) remove it? 
        $\bigcirc$ Definitely yes
        $\bigcirc$ Probably yes
        $\bigcirc$ Might or might not
        $\bigcirc$ Probably no
        $\bigcirc$ Definitely no
    \end{enumerate}
    
    \item Can you think of any other technique that can protect user deletions in presence of the malicious entity mentioned above? 
    (Explain in 1-3 sentences)
        
\end{enumerate}

\vspace{5em}
\subsection{Demographic Questions}
\hrule
\vspace{1em}

\noindent What is your age?\\ 
$\bigcirc$ Under 18
$\bigcirc$ 18 - 24
$\bigcirc$ 25 - 34
$\bigcirc$ 35 - 44
$\bigcirc$ 45 - 54
$\bigcirc$ 55 - 64
$\bigcirc$ 65 - 74
$\bigcirc$ 75 - 84
$\bigcirc$ 85 or older
$\bigcirc$ Prefer not to answer\\

\noindent Which gender do you identify most with?\\  
$\bigcirc$ Male
$\bigcirc$ Female
$\bigcirc$ Other
$\bigcirc$ Prefer not to answer\\

\noindent Please specify your ethnicity.\\ 
$\bigcirc$ African American
$\bigcirc$ Asian
$\bigcirc$ Hispanic or Latino
$\bigcirc$ Native American
$\bigcirc$ Middle Eastern
$\bigcirc$ White or Caucasian
$\bigcirc$ Multiple races
$\bigcirc$ Others
$\bigcirc$ Prefer not to answer\\

\noindent What is the highest level of school you have completed or the highest degree you have received?\\
$\bigcirc$ Less than high school degree
$\bigcirc$ High school degree or equivalent (e.g., GED)
$\bigcirc$ Some college but no degree
$\bigcirc$ Associate degree
$\bigcirc$ Bachelor degree
$\bigcirc$ Graduate degree
$\bigcirc$ Prefer not to answer\\

\noindent What is your marital status?  \\
$\bigcirc$ Single, never married
$\bigcirc$ Married or domestic partnership
$\bigcirc$ Widowed
$\bigcirc$ Divorced
$\bigcirc$ Separated
$\bigcirc$ Prefer not to answer\\

\noindent Which of the following categories best describes your employment status?\\
$\bigcirc$ Full-time employment
$\bigcirc$ Part-time employment
$\bigcirc$ Unemployed
$\bigcirc$ Full time uncompensated (e.g., homemaker, volunteer)
$\bigcirc$ Student
$\bigcirc$ Retired
$\bigcirc$ Other
$\bigcirc$ Prefer not to answer\\

\noindent Which annual income group does your household fall under?\\
$\bigcirc$ \$0 - \$9,999
$\bigcirc$ \$10,000 - \$19,999
$\bigcirc$ \$20,000 - \$29,999
$\bigcirc$ \$30,000 - \$39,999
$\bigcirc$ \$40,000 - \$49,999
$\bigcirc$ \$50,000 - \$59,999
$\bigcirc$ \$60,000 - \$69,999
$\bigcirc$ \$70,000 - \$79,999
$\bigcirc$ \$80,000 - \$89,999
$\bigcirc$ \$90,000 - \$99,999
$\bigcirc$ \$100,000 or more
$\bigcirc$ Prefer not to answer\\

\noindent Do you currently have a job (or previously worked) in computer science, information technology, or some other technical field? Or, if you are a student, do you study one of these topics in your degree program?\\
$\bigcirc$ Yes
$\bigcirc$ No
$\bigcirc$ Prefer not to answer\\

\end{document}